\documentclass[twocolumn,superscriptaddress,floatfix,preprintnumbers,prd,nofootinbib]{revtex4-2}
\usepackage[colorlinks=true,breaklinks=true]{hyperref}
\usepackage[utf8]{inputenc}
\hypersetup{urlcolor=Blue,
	    citecolor=Blue,
	    linkcolor=Blue}
\usepackage{url}
\usepackage{mathtools}
\usepackage[dvipsnames]{xcolor}

\usepackage{orcidlink}

\newcommand{\unit}[1]{\,\mathrm{#1}}
\newcommand{\unitspace}{\ensuremath{\mskip\thinmuskip}}
\newcommand{\Msol}{\ensuremath{\mathrm{M_{\odot}}}}
\newcommand{\gccm}{\ensuremath{{\mathrm{g}\unitspace\mathrm{cm}^{-3}}}}
\newcommand{\derive}[0]{\mathrm{d}}

\newcommand{\nue}{\ensuremath{\nu_{e}}}
\newcommand{\nuebar}{\ensuremath{\bar{\nu}_{e}}}
\newcommand{\nux}{\ensuremath{\nu_{x}}}
\newcommand{\EtoX}{\ensuremath{\nue,\nuebar \to \nux} }
\newcommand{\XtoE}{\ensuremath{\nux \to\nue,\nuebar} }

\long\def\exclude#1{}

\bibpunct{[}{]}{,}{n}{}{,}

\begin{document}

\title{Fast Neutrino Flavor Conversion in Core-Collapse Supernovae:\\ A Parametric Study in 1D Models}

\author{Jakob Ehring \orcidlink{0000-0003-2912-9978}}
\affiliation{Max-Planck-Institut f\"ur Physik (Werner-Heisenberg-Institut), F\"ohringer Ring 6, D-80805 M\"unchen, Germany}
\affiliation{Max-Planck-Institut f\"ur Astrophysik, Karl-Schwarzschild-Str.~1, D-85748 Garching, Germany}
\affiliation{Physik Department, Technische Universit\"at M\"unchen, James-Franck-Str.~1, D-85748 Garching, Germany}

\author{Sajad Abbar \orcidlink{0000-0001-8276-997X}}
\affiliation{Max-Planck-Institut f\"ur Physik (Werner-Heisenberg-Institut), F\"ohringer Ring 6, D-80805 M\"unchen, Germany}

\author{Hans-Thomas Janka \orcidlink{0000-0002-0831-3330}}
\affiliation{Max-Planck-Institut f\"ur Astrophysik, Karl-Schwarzschild-Str.~1, D-85748 Garching, Germany}

\author{Georg Raffelt \orcidlink{0000-0002-0199-9560}}
\affiliation{Max-Planck-Institut f\"ur Physik (Werner-Heisenberg-Institut), F\"ohringer Ring 6, D-80805 M\"unchen, Germany}

\author{Irene Tamborra \orcidlink{0000-0001-7449-104X}}
\affiliation{Niels Bohr International Academy \& DARK, Niels Bohr Institute, University of Copenhagen, Blegdamsvej 17, DK-2100 Copenhagen, Denmark}

\date{02 April 2023}


\begin{abstract}
We explore the impact of small-scale flavor conversions of neutrinos, the so-called fast flavor conversions (FFCs), on the dynamical evolution and neutrino emission of core-collapse supernovae (CCSNe). In order to do that, we implement FFCs in the spherically symmetric (1D) CCSN simulations of a 20\,M$_\odot$ progenitor model parametrically, assuming that FFCs happen at densities lower than a systematically varied threshold value and lead to an immediate flavor equilibrium consistent with lepton number conservation. We find that besides hardening the $\nu_{e}$ and $\bar{\nu}_{e}$ spectra, which helps the expansion of the shock by enhanced postshock heating, FFCs can cause significant, nontrivial modifications of the energy transport in the SN environment via increasing the $\nu_{\mu, \tau}$ luminosities. In our non-exploding models this results in extra cooling of the layers around the neutrinospheres, which triggers a faster contraction of the proto-neutron star and hence, in our 1D models, hampers the CCSN explosion. Although our study is limited by the 1D nature of our simulations, it provides valuable insights into how neutrino flavor conversions in the deepest CCSN regions can impact the neutrino release and the corresponding response of the stellar medium.
\end{abstract}
\maketitle


\section{Introduction and Motivation}
\label{sec:introduction}

Core-collapse supernova (CCSN) explosions are among the most extreme astrophysical phenomena in the universe.
Apart from being exceptionally luminous in electromagnetic radiation, they are also a site of intense neutrino production \citep[e.g.,][]{Janka:2012wk,Janka2017Handbookb,Mirizzi:2015eza,Burrows:2020qrp,Mueller:2019upo,Mezzacappa:2020oyq}.
Within roughly 10\,s, $\mathcal{O}(10^{53})$~erg of gravitational binding energy are converted to $\mathcal{O}(10^{58})$ neutrinos with a mean escape energy of about $10\,\unit{MeV}$.
Understanding the processes that take place in the deep interior of CCSNe requires a solid understanding of the behavior of neutrinos.
Conversely, CCSNe provide us with a unique opportunity to test predictions about the nature of neutrinos otherwise not accessible.

Numerous studies have already refined our knowledge of different aspects of the interplay between neutrinos and the stellar medium, where the neutrino flavor plays an important role.
Because the smaller electron mass permits a much larger abundance of electrons compared to muons and taus, $\nue$ and $\nuebar$ interact with the supernova (SN) medium more strongly than the heavy-lepton neutrinos.
This obviously necessitates a distinction of the different neutrino flavors in such environments.
Yet, although it has been known for a long time that neutrinos are able to undergo flavor conversion during their propagation due to their quantum mechanical nature, state-of-the-art CCSN simulations assume that neutrinos retain their flavor identity after their production.
In this study, we suspend this assumption and explore the impact of neutrino flavor conversion on the core-collapse dynamics, proto-neutron star (PNS) formation, and the neutrino transport and emission from collapsing stars.

An intriguing feature of the neutrino flavor evolution in CCSNe is connected to the extremely high neutrino number densities in these environments, which have the consequence that neutrino-neutrino interactions play an important role in the neutrino flavor evolution.
Neutrino-flavor conversions in CCSNe have turned out to be a very rich, nonlinear collective phenomenon~\cite{Pastor:2002we,duan:2006an,Fogli:2007bk,duan:2010bg, Duan:2009cd, Mirizzi:2015eza, Volpe:2023met}.
In particular, neutrinos can experience the so-called \emph{fast} flavor conversions (FFCs), in which neutrinos and antineutrinos can undergo pairwise changes of their flavor, provided that their propagation is not entirely parallel~\citep[e.g.,][]{Sawyer:2005jk, Sawyer:2015dsa,Chakraborty:2016lct, Izaguirre:2016gsx,Capozzi:2017gqd,Abbar:2018beu,Capozzi:2018clo, Martin:2019gxb, Capozzi:2019lso, Johns:2019izj, Martin:2021xyl, Tamborra:2020cul, Sigl:2021tmj, Kato:2021cjf, Morinaga:2021vmc, Nagakura:2021hyb, Sasaki:2021zld, Padilla-Gay:2021haz, Abbar:2020qpi, Capozzi:2020syn, DelfanAzari:2019epo, Harada:2021ata, Padilla-Gay:2022wck, Capozzi:2022dtr, Zaizen:2022cik, Shalgar:2022rjj, Kato:2022vsu, Zaizen:2022cik, Bhattacharyya:2020jpj, Wu:2021uvt, Richers:2021nbx, Richers:2021xtf, Dasgupta:2021gfs, Nagakura:2022kic}.
One can prove that crossings of the angular distributions of $\nue$ and $\nuebar$ are necessary and sufficient conditions for FFCs to occur (in a situation where $n_{\nu_{\mu, \tau}} = n_{\bar\nu_{\mu, \tau}}$), i.e., zero crossings of the so-called electron neutrino lepton number (ELN) angular distribution should exist~\cite{Morinaga:2021vmc}.
A characteristic feature of FFCs is the fact that the spatial scale of such conversions can be as short as centimeters in the deepest regions of the SN core, as the relevant length scale depends on the neutrino number densities, $n_\nu$.
This should be compared with the traditional \emph{slow} modes, where flavor conversions occur on scales determined by the neutrino vacuum frequency, corresponding to length scales of the order of a few kilometers for typical SN neutrinos.

It has been demonstrated that ELN crossings can occur in three different SN regions:
(i) In the convection layer inside the PNS, where the neutrino gas becomes nondegenerate due to the progressing deleptonization, which is accelerated by convective lepton-number transport~\cite{DelfanAzari:2019tez,Abbar:2019zoq,Glas:2019ijo};
(ii) within the neutrino decoupling layer and the near-surface region of the PNS, where the neutrino-antineutrino asymmetry parameter, $n_{\bar\nu_e}/n_{\nu_e}$, is relatively close to unity~\cite{Abbar:2018shq,Abbar:2020qpi,Nagakura:2019sig,Shalgar:2019kzy};
and (iii) at larger distances from the SN core, both in the preshock and postshock regions~\cite{Morinaga:2019wsv, Capozzi:2020syn}.
Perhaps the most interesting regions where FFCs could happen are the neutrino decoupling layer and the convective shell in the PNS, although the occurrence of ELN crossings does not guarantee the development of significant FFC with implications on the source physics.
Deeper inside the PNS interior, FFCs are not expected to cause significant flavor changes as long as the electron neutrinos are highly degenerate (though this condition abates at later post-bounce times because of the continuous deleptonization).
Moreover, FFCs occurring at large distances from the SN core could be suppressed by the shape of the unstable flavor states~\cite{Abbar:2021lmm}.

Although many studies have explored the physics linked to FFCs in a dense neutrino gas, the consequences of such conversions for the evolution of CCSNe are not clear yet and our understanding of the physics of FFC is still very preliminary~\cite{Tamborra:2020cul,Richers:2022zug}.
Given the complex, nonlinear feedback effects that play a role in the coupled neutrino-hydrodynamics problem, any attempt to pin down the dynamical impact of FFCs must include their self-consistent implementation in CCSN simulations.
However, a full-fledged quantum kinetic treatment of neutrinos combined with the solution of the hydrodynamics of the SN plasma is currently (and will remain so in the near future) an unfeasible task from the computational point of view.
One obstacle is connected to the fact that the stiff and non-linear nature of the problem hampers the derivation of full-scale, multi-dimensional solutions of the quantum kinetic equations of neutrinos as well as a robust assessment of the amount of flavor conversion that can be achieved.
Another difficulty arises because the FFC scales in space and time are expected to be orders of magnitude smaller than any other dynamical scales of relevance in the SN environment, which have to be resolved by the discrete spatial zoning and time stepping applied in the numerical calculations.
This vast separation of scales significantly impedes the self-consistent implementation of FFCs in CCSN models, but it can also serve as a motivation for a schematic description of FFC effects in the simulations (e.g., \cite{Li:2021vqj, Just:2022flt}).
A schematic approach, not based on rigorous solutions of the quantum kinetic problem, must make assumptions for when and where FFCs occur, how these flavor conversions develop in space and time, and which final state of the neutrino distributions is reached by the flavor conversion.

In this paper we report the results of an investigation where, for the first time, we explore the impact of FFCs in spherically symmetric (1D) neutrino-hydrodynamic CCSN simulations.
We employ a schematic treatment in which we use a simple density criterion to systematically vary the spatial region where FFCs are assumed to take place.
In this region, during our entire simulations, we apply an effective, lepton-number conserving flavor-equilibration scheme that considers flavor conversion to happen instantaneously, i.e., on time scales shorter than the time step for hydrodynamics and neutrino transport, and on short length scales smaller than the numerical grid-resolution scale.
Moreover, we maximize the possible impact of FFCs by assuming full equilibration under the constraint of lepton number conservation for all flavors (in particular also of electron lepton number).
1D models provide the possibility to disentangle the consequences of the FFCs from the additional complexity of multi-dimensional effects.
Irrespective of the fact that ELN crossings have not been observed in the neutrino decoupling region of 1D models~\cite{Tamborra:2017ubu}, we apply our prescription of flavor conversions in different regions where the matter density $\rho$ drops below a threshold value $\rho_{\mathrm{c}}$, i.e., our criterion for the occurrence of FFCs is $\rho < \rho_{\mathrm{c}}$ with $\rho_{\mathrm{c}}$ running from $10^9$\,g\,cm$^{-3}$ to $10^{14}$\,g\,cm$^{-3}$.
This allows us to explore the influence of FFCs happening in different regions of the SN core.
In general, our results show that FFC can modify the heating as well as cooling of the CCSN environment in a non-trivial manner.
In particular, this leads to case-dependent, short phases of stronger shock expansion as well as faster PNS contraction and more extreme shock recession during the long-time post-bounce evolution in our non-exploding 1D models.
Multi-dimensional simulations are required to infer the implications of such effects for the CCSN mechanism.

At the time of submission, a preprint appeared~\cite{Nagakura:2023mhr} along the lines of our study, but solving the quantum kinetic equations to determine the effects of FFCs in a 1D modified hydrostatic and fixed background from a simulation and without feedback of the flavor conversion physics on the CCSN dynamics.
The emerging steady-state solution for the neutrino flux leads the author to conclusions partially in line with the results from our hydrodynamic simulations, which assume maximum equipartition under the constraint of ELN conservation and account for the feedback of FFCs on the time-dependent evolution.

Our paper is structured as follows.
In Section~\ref{sec:Setup} we describe the computational setup used for our simulations including the numerical code and the implementation of FFCs.
In Section~\ref{sec:Results} we discuss our results, and in Section~\ref{sec:conclusion} we finish with a summary and conclusions.

\section{Setup for the simulation}\label{sec:Setup}

For our simulations we use the well established \textsc{Aenus-Alcar} code \citep{Obergaulinger2008PhDT,Just:2015fda,Just:2018djz}, whose major elements are summarized in Section~\ref{sec:numerical_code}.
We extended the code to include an effective treatment of FFCs in a parametrized way.
In this treatment, we enforce equipartition of neutrinos and antineutrinos in the flavor states at each time step with the constraint that all individual lepton numbers, in particular the electron lepton number, are conserved, as discussed in Section~\ref{sec:implementation_ffc}.
We use the 20\,M$_\odot$ progenitor model, denoted by S20, from \citep{2007PhR...442..269W} for all our simulations.
As expected, the 1D simulation with no flavor conversions (NFC) does not yield an explosion, and we find that also none of our simulations with FFCs produces an explosion in the simulated periods of at least 0.5\,s of post-bounce evolution.

\subsection{Numerical Code}\label{sec:numerical_code}

Our 1D simulations are performed with the spherically symmetric version of \textsc{Alcar}~\cite{Just:2015fda}.
The code solves the hydrodynamics equations of the stellar fluid (Eqs.~1a--d in Ref.~\cite{Just:2018djz}), which are closed by the microphysical SFHo equation of state (EoS)~\citep{Steiner:2012rk} extended to a minimum temperature of $10^{-3}\unit{MeV}$.
The EoS treats the nuclear composition in nuclear statistical equilibrium.
This applies accurately in the postshock layer of our non-exploding models, and it is a reasonable approximation in the infall regions ahead of the stalled shock, where it accounts for the budget of nuclear energy overall appropriately, although it does not follow the detailed nuclear composition correctly.

\textsc{Alcar} treats the transport for three neutrino species, electron neutrinos $\nue$, electron antineutrinos $\nuebar$, and heavy-lepton neutrinos $\nux$, which represent $\nu_{\mu}$, $\nu_{\tau}$, $\bar{\nu}_{\mu}$, and $\bar{\nu}_{\tau}$, whose interaction with the stellar medium is very similar (for differences connected to physics that requires the separate tracking of six instead of three neutrino species, see Refs.~\cite{Mirizzi:2015eza,Bollig:2017lki,Bollig:2018}).
The neutrino transport is solved via a two-moment scheme (Eqs.~3a--b in Ref.~\cite{Just:2018djz}), and the moment equations for neutrino energy density and flux density are closed by using the Minerbo closure, where the 2nd and 3rd-order angular moments of the neutrino distributions are obtained algebraically from the 0th and 1st-order ones.

In our CCSN calculations the transport of all types of neutrinos is taken into account from the beginning of the simulations, i.e., also during the collapse phase until core bounce.
The following neutrino interactions are included in the transport and yield source terms for energy, momentum, and electron lepton number in the hydrodynamics equations of the stellar medium:
\begin{enumerate}
 \item Charged-current ($\beta$) processes and neutral-current scattering on nucleons (as in Refs.~\cite{1985ApJS...58..771B,1993ApJ...405..637M,2002PhRvD..65d3001H})
 \item $\beta$-processes of nuclei~\citep{1985ApJS...58..771B,1993ApJ...405..637M},
 \item coherent scattering with nuclei~\citep{1997PhRvD..56.7529B,1997PhRvD..55.4577H},
 \item scattering with electrons and positrons (see, e.g., Refs.~\cite{1977ApJ...217..565Y,1985ApJS...58..771B,1994ApJ...433..247C}),
\item thermal processes for heavy-lepton neutrino-anti\-neutrino pairs:
     \begin{enumerate}
         \item electron-positron annihilation \citep[][]{1998A&AS..129..343P},
         \item nucleon-nucleon bremsstrahlung \citep[][]{1998ApJ...507..339H}.
     \end{enumerate}
\end{enumerate}

Both the hydrodynamics and transport solvers use a Godunov-type finite-volume scheme in spherical polar coordinates with the time integration being done by an explicit, second-order Runge-Kutta method, where the time step is constrained by the Courant-Friedrichs-Lewy (CFL) condition \citep{1928MatAn.100...32C}.
We use a logarithmically spaced radial grid of 640 zones up to an open boundary at 10,000\,km for both hydrodynamics and transport and employ a logarithmically spaced energy grid of 15 energy bins up to 400\,MeV (outer boundary of the energy grid).
Some source terms for the interaction processes of the neutrinos are treated time-implicitly in the transport equations (for detailed explanations, see Appendix A of Ref.~\cite{Just:2018djz}).
The source terms for gravity in spherical symmetry are computed by a Newtonian potential, modified with general relativistic corrections according to case~A of Ref.~\cite{Marek:2005if}.

The 0th angular moment (the neutrino energy density, $E_{\nu_{\alpha}}$) and 1st angular moment (the neutrino energy flux density, $\boldsymbol{F}_{\nu_{\alpha}}$) of the phase-space distribution of neutrino species $\nu_\alpha$, whose time evolution is computed by the two-moment solver, are defined as
\begin{equation}
    \{cE_{\nu_{\alpha}}, \boldsymbol{F}_{\nu_{\alpha}}\} = \int \derive{\Omega_{\boldsymbol{n}}}\ \mathcal{I_{\nu_{\alpha}}} \{1,\boldsymbol{n}\}\,,
\end{equation}
where $\mathcal{I_{\nu_{\alpha}}}$ and $\boldsymbol{n}$ are the specific intensity (i.e., for a given neutrino energy $\epsilon$) and the unit vector that points in the direction of the neutrino momentum, respectively.
$\mathcal{I}_{\nu_{\alpha}}$ is derived from the phase-space distribution function $\mathcal{F_{\nu_{\alpha}}}$ as
\begin{equation}
    \mathcal{I_{\nu_{\alpha}}} = g_\alpha\left(\frac{\epsilon}{hc}\right)^{\! 3}c\mathcal{F_{\nu_{\alpha}}}\,,
\end{equation}
where $c$ is the speed of light, $h$ the Planck constant, $g_{\alpha}$ the statistical weight of the neutrino or antineutrino ($g_{\alpha} = 1$ for one species), and $\epsilon = \lvert\boldsymbol{p}\rvert c$ is the neutrino energy with momentum $\boldsymbol{p}$.
The number of neutrinos of species $\nu_\alpha$ in a phase-space volume $\derive{x}^3\derive{p}^3$ can then be found as
\begin{equation}
    \derive{N_{\nu_{\alpha}}} = \frac{g_{\alpha}}{h^3}\,\mathcal{F_{\nu_{\alpha}}}\derive{x}^3\derive{p}^3\,.
\end{equation}
For more details, we refer to Refs.~\citep{Just:2015fda,Just:2018djz}, whose notation we followed in this section.

\subsection{Fast Flavor Conversion --- Motivation of Schematic Treatment}

Our understanding of the physics of FFC is still very preliminary \citep[e.g.,][]{Tamborra:2020cul,Richers:2022zug}.
As mentioned before, the stiff and non-linear nature of the problem impedes the development of full-scale, multi-dimensional solutions of the quantum kinetic equations of neutrinos as well as a robust assessment of the amount of flavor conversion that one should expect.
Besides intense work to determine the locations where FFCs may take place in 1D and multi-D CCSN models (see Section~\ref{sec:introduction}), a large number of studies has also been devoted to make progress on understanding how the flavor transformation proceeds and which flavor mix it leads to.
In order to motivate the schematic description of FFCs adopted in our work, we refer to insights obtained from the corresponding wealth of recent, direct calculations of the quantum kinetic problem for specifically defined, mostly idealized conditions.

Based on results from toy models that used setups with imposed initial angular distributions, periodic boundary conditions and neutrino inhomogeneities, it was pointed out that FFCs might lead to a sort of flavor equilibrium in a multi-dimensional neutrino gas \cite{Bhattacharyya:2020jpj,Wu:2021uvt,Richers:2021nbx,Richers:2021xtf,Nagakura:2022kic,Richers:2022bkd}.
However, this possibility must still be considered as controversial in view of Refs.\ \cite{Shalgar:2022rjj,Shalgar:2022lvv,Nagakura:2022kic,Nagakura:2022qko}, which aimed at simulating the evolution of the neutrino field on large scales in the presence of FFCs by gradually evolving the neutrino distributions and their flavor when neutrinos decouple from the matter background.
These works suggest that flavor equilibration may not be the general outcome but rather one possible result of a subtle interplay between the shape of the angular distributions, the strength of collisions (i.e., neutrino interactions with particles of the stellar medium), and neutrino propagation (see also Ref.~\cite{Shalgar:2019qwg} for an earlier study reaching similar conclusions).

The interplay between collisions and FFCs has triggered particular interest because of a puzzling enhancement or suppression of flavor conversions, depending on the shape of the ELN distributions, see, e.g., Refs.~\cite{Shalgar:2020wcx,Kato:2022vsu,Johns:2022bmu,Hansen:2022xza,Lin:2022dek,Sasaki:2021zld}.
Intriguingly, in addition to the impact that collisions could have on FFCs, collision induced flavor instability \cite{Johns:2021qby, Johns:2022yqy, Xiong:2022vsy, Xiong:2022zqz, Lin:2022dek} could be another way to trigger FFCs in the presence of a large asymmetry between the collision terms of $\nu_e$ and $\bar\nu_e$ \cite{Johns:2021qby,Xiong:2022vsy,Xiong:2022zqz,Johns:2022yqy,Liu:2023vba,Padilla-Gay:2022wck}.
Provided such a causal link exists, it is conceivable that in hydrodynamic CCSN simulations not only FFCs could be approximated by a schematic method as applied by us (and described in detail in Section~\ref{sec:implementation_ffc}), but also the impact of collision induced flavor instability or any other flavor conversion phenomenon could be effectively treated in a similar way.
This would only require that the flavor conversions occur on sufficiently short length and time scales and conserve ELN number.

Collision induced conversions actually develop on characteristic scales that are determined by the collision term in the transport equation and thus by the spatial scales of relevance for the neutrino transport.
These spatial scales can be much bigger than the local radius or the numerical grid cell size in regions where the neutrino mean free paths are large.
However, if collision induced conversions trigger FFCs, the fast conversion process that kicks in would again justify our simplified parametrization that is supposed to mimic FFCs in our CCSN simulations, assuming flavor mixing happens on time and length scales shorter than the numerical time stepping and the spatial grid resolution.

\begin{figure}
    \includegraphics{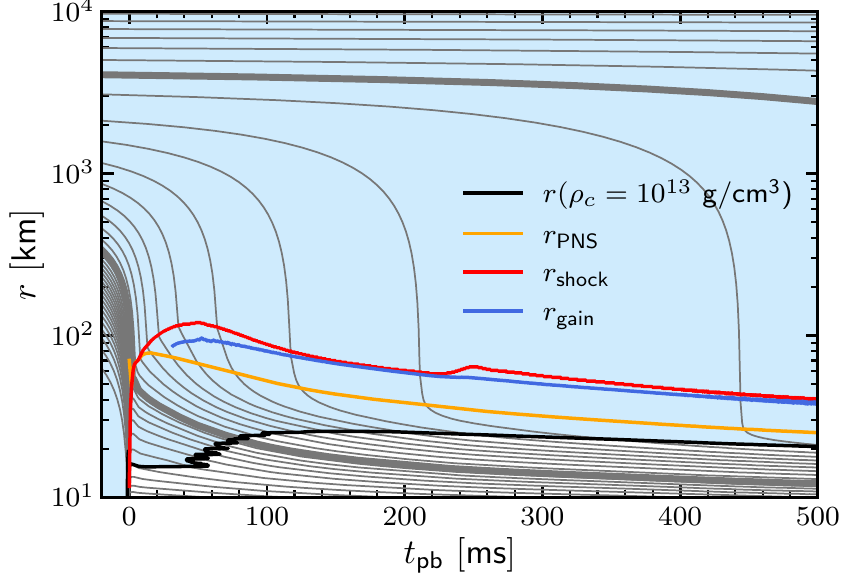}
    \caption{Mass-shell diagram of model M-1e13.
    The gray lines correspond to selected mass shells, i.e., to the time-dependent radial locations of chosen, fixed values of the enclosed mass.
    The thick gray lines mark the positions of 1\,M$_\odot$ and 2\,M$_\odot$, respectively, and the thin gray lines are spaced with intervals of 0.05\,M$_\odot$ interior to 1\,M$_\odot$ and with 0.1\,M$_\odot$ outside.
    The PNS radius (defined at a density of $\rho=10^{11}$\,g\,cm$^{-3}$), shock radius, and gain radius are indicated by orange, red, and blue lines, respectively.
    The black line marks the isodensity contour for a value of $\rho=10^{13}$\,g\,cm$^{-3}$ and separates the regions where we apply (outside, light-blue shaded) or do not apply (inside, unshaded) our flavor equilibration scheme.}
    \label{fig:massshell}
\end{figure}

\subsection{Implementation of FFC}\label{sec:implementation_ffc}

Because of the technical challenges of rigorous solutions discussed above, any attempt to implement FFCs in CCSN simulations must currently be done schematically, given the fact that the FFC scales are expected to be much smaller than any other scales of relevance for the SN dynamics.
In a schematic method, one can therefore assume that FFCs lead to spontaneous and instantaneous pair-wise changes in the abundances of neutrinos and antineutrinos of different flavors.
This means that in our numerical treatment we apply our flavor-conversion prescription in each time step and in each grid cell identified to be subject to FFCs according to a simple criterion.
Although it is not yet clear whether FFCs lead to flavor equilibrium, we assume full flavor equilibration of the neutrinos under the constraint of lepton-number conservation, in particular also of electron lepton number.
We thus aim at maximizing the impact of FFCs and therefore at assessing the consequences of flavor conversions on the CCSN physics in the most extreme FFC scenario.

There are two key components of any such schematic implementation of FFCs in CCSN simulations.
First, one needs a strategy to capture the SN zones where ELN crossings exist.
Second, a prescription for the equilibrium state that results from FFCs is required.

Considering that ELN crossings have not been observed in the neutrino decoupling region of 1D CCSN models, as mentioned before, we apply a simple recipe based on the matter density in order to perform a parametric study of the impact of FFCs occurring in different regions of the SN core.
To be specific, we introduce a threshold density, $\rho_\mathrm{c}$, below which FFCs are assumed to happen, i.e., for $\rho < \rho_\mathrm{c}$.
Usually (but not always) this means that our schematic prescription of FFCs is applied in a region $r > r_\mathrm{c}$ exterior to the radius $r_\mathrm{c}$ where $\rho(r_\mathrm{c}) = \rho_\mathrm{c}$.
While the value of the threshold density $\rho_\mathrm{c}$ is constant (i.e., time independent) and specific for each of our CCSN simulations, the spatial volume where FFCs are assumed to happen evolves with time, which means that the radius (or radii) that deliminates the region of FFCs evolves with time, i.e., $r_\mathrm{c} = r_\mathrm{c}(t)$; see the black line and blue shaded region in Figure~\ref{fig:massshell} for an exemplary case of one of our CCSN simulations.
Beginning with a CCSN model where the density $\rho_\mathrm{c}$ is fixed to $10^9\,\gccm$, which corresponds to a region exterior to the stalled shock front in all of our non-exploding models, we fix $\rho_\mathrm{c}$ in the other simulations to values higher by factors of $10$ in steps up to $10^{14}\,\gccm$, which is a density well inside the convectively unstable layer of the PNS.
Thus we scan different values of $\rho_\mathrm{c}$ and are able to develop an understanding of possible nonlinear feedback effects of neutrino flavor conversions in all potentially affected regions on the SN dynamics and neutrino emission.
We also test intermediate values for the threshold densities, i.e., $\rho_\mathrm{c} = 10^{n+0.5}$\,g\,cm$^{-3}$ with $n = 9, 10,....., 13$, but cannot find any new features beyond those reported in Section~\ref{sec:Results}.

Concerning the prescription for equilibration, one needs to define the equilibrium values of the 0th and 1st neutrino moments in the M1 scheme as functions of their initial values.
Here we assume that the equilibrium develops separately in each energy bin.
This treatment ensures energy conservation and can be justified by the fact that forward scattering does not change neutrino energies and directions.

\begin{figure}
    \includegraphics{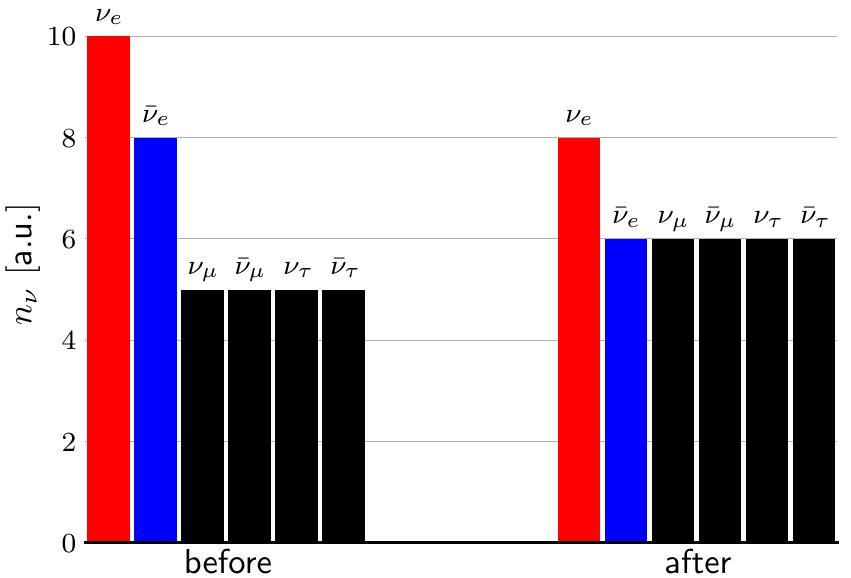}\\\vspace{15pt}
    \includegraphics{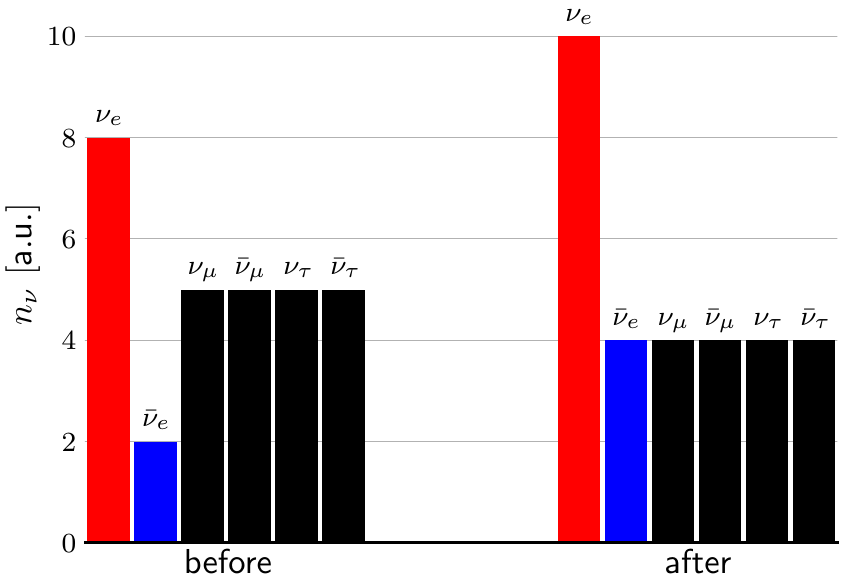}
    \caption{Schematic representation of the equilibrium prescription introduced in Equation~(\ref{Eq:equilibrium1}) for $L_{e,q}>0$.
    The neutrino gas reaches an equipartition between $\nux$ and $\nuebar$, which are less abundant than $\nue$, with $L_{e,q}$ remaining unchanged.
    The upper panel is for \EtoX conversions, the lower panel for \XtoE conversions.}
    \label{fig:schematic}
\end{figure}

Let us first consider the 0th moment, i.e., the specific neutrino number or energy densities.
With our numerical discretization, the number density of neutrinos of species $\nu_\alpha$ in the $q$th energy bin, $n_{\nu_{\alpha},q}$, is related the neutrino energy density $E_{\nu_{\alpha},q}$ in that bin by
\begin{equation}
    n_{\nu_{\alpha},q} = \frac{E_{\nu_{\alpha},q}}{\epsilon_{q}}\,,
\end{equation}
with $\epsilon_{q}$ being the mean energy of the $q$th energy bin.
In the following, we will use a notation where the unprimed variables refer to the initial values of the physical quantities (before applying our flavor conversion prescription), and the primed variables refer to their flavor equilibrated values (after applying our prescription).
Our prescription for the final number densities is based on four requirements:
\begin{itemize}
    \item As discussed above, FFCs cannot change the total numbers of neutrinos and antineutrinos in each energy bin.
    Hence, our prescription conserves the summed numbers of neutrinos and antineutrinos separately:
    \begin{align}
        &n_{\nue,q}    + 2 n_{\nux,q} = n'_{\nue,q}    + 2 n'_{\nux,q}\,,\nonumber\\
        &n_{\nuebar,q} + 2 n_{\nux,q} = n'_{\nuebar,q} + 2 n'_{\nux,q}\,.
    \end{align}

    \item FFCs themselves cannot change the neutrino ELN, $L_{e,q} = n_{\nue,q} - n_{\nuebar,q}$, which is therefore conserved by our prescription:
    \begin{equation}
        L_{e,q} = n_{\nue,q} - n_{\nuebar,q} = n'_{\nue,q} - n'_{\nuebar,q} = L'_{e,q} \,.
    \end{equation}

    \item Any physical prescription should respect the Pauli blocking.
    This can be an important issue in the regions with high electron degeneracy, where also the electron neutrino number densities can approach the Pauli limit,
    \begin{equation}
        n_{\mathsf{Pauli},q} = \frac{4\pi \epsilon_q^2 \Delta\epsilon_q}{(h c)^3}\,,
    \end{equation}
    with $\Delta\epsilon_{q}$ being the width of the $q$th energy bin, and \XtoE conversions should not overpopulate the bins in order to avoid the violation of Pauli blocking.
    This constraint is particularly relevant for the lower energy bins with $\epsilon_q \lesssim k_\mathrm{B}T$, when $T$ is the temperature of the fluid and $k_\mathrm{B}$ the Boltzmann constant.

    \item Although the physical quantities of the FFC equilibrium state can in general be nonlinear and complicated functions of the initial values of these quantities \citep[e.g.,][for calculations that found flavor equilibration]{Richers:2022bkd, Richers:2021xtf, Richers:2021nbx, Wu:2021uvt, Bhattacharyya:2020jpj}, we assume here that the final quantities are linear functions of the initial ones.
    Such a simple equilibrium prescription can significantly constrain the choices for the suitable expressions of the equilibrium values.
    In particular, our prescription is motivated by the flavor equilibrium observed for slow modes and small values of $\nu_e$-$\bar\nu_e$ asymmetries in 1D neutrino gases~\cite{Esteban-Pretel:2007jwl}.
    Specifically, we assume an equilibrium state where equipartition occurs between $\nu_x$ and $\nue$ or $\nux$ and $\nuebar$, depending on whether electron neutrinos or antineutrinos, respectively, are less abundant before applying our prescription (see Figure~\ref{fig:schematic}).
    This implies that the flavor-equilibrium number density $n_{\mathrm{eq},q}$ in the $q$th energy bin of the subdominant species is given by
    \begin{equation}
        n_{\mathrm{eq},q} =
        \begin{cases}
            n'_{\nuebar,q} = n'_{\nux,q} = \frac{1}{3}(n_{\nuebar,q} + 2 n_{\nux,q})\,,    & \text{if } L_{e,q} > 0\,, \\
            n'_{\nue,q}    = n'_{\nux,q} = \frac{1}{3}(n_{\nue,q} + 2 n_{\nux,q})\,,       & \text{if } L_{e,q} \leq 0\,.
        \end{cases}
    \end{equation}
\end{itemize}

Applying these conditions, the flavor-equilibrium values of the neutrino and antineutrino number densities can be determined as
\begin{subequations}\label{Eq:equilibrium1}
    \begin{align}
        n_{\nue,q}'    & = n_{\mathrm{eq},q}  + \max(0,L_{e,q})\,, \\
        n_{\nuebar,q}' & = n_{\mathrm{eq},q}  + \max(0,-L_{e,q})\,,\\
        n_{\nux,q}'    & =  n_{\mathrm{eq},q}\,,
    \end{align}
\end{subequations}
for $n_{\nu,q}' < n_{\mathsf{Pauli},q}$.
One can see that this prescription puts the neutrino gas into equipartition up to the constraint of ELN conservation.
This is illustrated schematically in Figure~\ref{fig:schematic}.
It should be noted that in a realistic situation the induced neutrino flavor conversions might be weaker than assumed in our prescription, which means that our equilibration recipe maximizes the conversion effects, provided the lepton numbers of all three flavors are conserved.

Our prescription can lead to an overpopulation of an energy bin, putting there more neutrinos than is compatible with the Pauli exclusion principle, i.e., $n_{\nu,q}' > n_{\mathsf{Pauli},q}$ could occur for $\nue$ or $\nuebar$ with number densities computed from Eq.~(\ref{Eq:equilibrium1}).
In this case, this equation must be replaced~by
\begin{subequations}\label{Eq:equilibrium2}
    \begin{align}
        n_{\nue,q}'    & = \mathrm{min}( n_{\mathsf{Pauli},q},\ n_{\mathsf{Pauli},q} + L_{e,q})\,, \\
        n_{\nuebar,q}' & = \mathrm{min}( n_{\mathsf{Pauli},q},\ n_{\mathsf{Pauli},q} - L_{e,q})\,, \\
        n_{\nux,q}'    & = \textstyle{\frac{1}{2}}(n_{\mathrm{max},q} - n_{\mathsf{Pauli},q})\,,
    \end{align}
\end{subequations}
where
\begin{equation}
        n_{\mathrm{max},q} =
        \begin{cases}
            2 n_{\nux,q} + n_{\nue,q}\,,    & \text{if } L_{e,q} > 0\,, \\
            2 n_{\nux,q} + n_{\nuebar,q}\,, & \text{if } L_{e,q} \leq 0
        \end{cases}
\end{equation}
is the total number density of neutrinos or antineutrinos in the $q$th energy bin, depending on which one is larger.\footnote{Note that if $\nue$ could violate Pauli blocking, i.e., if $n'_{\nue,q} > n_{\mathsf{Pauli},q}$ holds for $n'_{\nue,q}$ from Eq.~(\ref{Eq:equilibrium1}), then $n_{\nue,q}>n_{\nux,q}>n_{\nuebar,q}$ and $L_{e,q} > 0$ hold as well, and Eq.~(\ref{Eq:equilibrium2}) yields $n_{\nue,q}' = n_{\mathsf{Pauli},q}$ and $n_{\nuebar,q}' = n_{\mathsf{Pauli},q} - L_{e,q}$.
In contrast, in the case of Pauli blocking being relevant for $\nuebar$, i.e., if Eq.~(\ref{Eq:equilibrium1}) leads to $n'_{\nuebar,q} > n_{\mathsf{Pauli},q}$, then $n_{\nuebar,q}>n_{\nux,q}>n_{\nue,q}$ and $L_{e,q} < 0$ are fulfilled, and Eq.~(\ref{Eq:equilibrium2}) yields $n_{\nuebar,q}' = n_{\mathsf{Pauli},q}$ and $n_{\nue,q}' = n_{\mathsf{Pauli},q} + L_{e,q}$.}

An astute reader will notice that the above expression just means that flavor conversions are allowed up to the Pauli blocking limit.
Note that applying our recipe separately for all energy bins might lead to a situation where at a given location both of the aforementioned cases of Pauli blocking could be encountered, though in different energy bins.

So far we have only discussed the prescription of flavor equilibrium for the neutrino number (or energy) densities, i.e., for the 0th moments.
In order to construct a similar prescription for the first moments, we make use of another constraint:
\begin{itemize}
    \item Since FFC does not modify the neutrino momentum, our prescription shall conserve the sum of the 1st moments of all neutrino species in each energy bin separately, although in reality the momentum is conserved for neutrinos and antineutrinos individually.
    However, our neutrino transport scheme does not distinguish between $\nu_x$ and $\bar\nu_x$, which implies that we can only satisfy the conservation of the total momentum in the neutrino-antineutrino gas:
    \begin{equation}
        \boldsymbol{F}_{\nue,q} + \boldsymbol{F}_{\nuebar,q} + 4 \boldsymbol{F}_{\nux,q} =
        \boldsymbol{F}'_{\nue,q} + \boldsymbol{F}'_{\nuebar,q} + 4 \boldsymbol{F}'_{\nux,q}\,.
        \label{Eq:momcons}
    \end{equation}
\end{itemize}

Now let
\begin{equation}
    \eta_{\nu_\alpha,q} = \frac{n_{\nu_\alpha,q}'}{n_{\nu_{\alpha,q}}}
\end{equation}
be the ratio of neutrinos of species $\nu_\alpha$ in the $q$th bin after and before the flavor conversion.
If $\eta_{\nu_\alpha,q} < 1$, this ratio yields the fraction of neutrinos that retain their flavor, whereas $(1-\eta_{\nu_\alpha,q})$ is the fraction that changes the flavor.
One can easily verify that the following prescriptions conserve the total momentum of neutrinos according to Eq.~(\ref{Eq:momcons}).
In the regions where \EtoX (i.e., $\eta_{\nue,q}, \eta_{\nuebar,q}<1$) we use
\begin{subequations}\label{Eq:mom1}
    \begin{align}
        \boldsymbol{F}_{\nue,q}'    & = \eta_{\nue,q}    \boldsymbol{F}_{\nue,q}\,,                                                                                     \\
        \boldsymbol{F}_{\nuebar,q}' & = \eta_{\nuebar,q} \boldsymbol{F}_{\nuebar,q}\,,                                                                                  \\
        \boldsymbol{F}_{\nux,q}'    & = \boldsymbol{F}_{\nux,q} +
        \frac{(1-\eta_{\nue,q})\boldsymbol{F}_{\nue,q}+(1-\eta_{\nuebar,q})\boldsymbol{F}_{\nuebar,q}}{4}\,,
    \end{align}
\end{subequations}
and in the regions where \XtoE (i.e., $\eta_{\nux,q}<1$) we apply
\begin{subequations}\label{Eq:mom2}
    \begin{align}
        \boldsymbol{F}_{\nue,q}'    & = \boldsymbol{F}_{\nue,q}    + 2 (1-\eta_{\nux,q})\boldsymbol{F}_{\nux,q}\,, \\
        \boldsymbol{F}_{\nuebar,q}' & = \boldsymbol{F}_{\nuebar,q} + 2 (1-\eta_{\nux,q})\boldsymbol{F}_{\nux,q}\,, \\
        \boldsymbol{F}_{\nux,q}'    & = \eta_{\nux,q} \boldsymbol{F}_{\nux,q}\,.
    \end{align}
\end{subequations}

It should be kept in mind that the prescription given by Eqs.~(\ref{Eq:mom1}) and (\ref{Eq:mom2}) is not unique in conserving the neutrino momentum.
However, it is physically intuitive, because it can be justified on the basis of two simple facts.
First, the momentum is effectively transferred from the original species to the one it is converted to.
Second, considering our relations of initial and final neutrino quantities, one can see that the flux factor ${f} = |\boldsymbol{F}|/cE$ remains constant during the flavor conversion in each energy bin $q$ of those species that effectively get converted to the other ones, i.e.,
    \begin{equation}
        {f}_{\nu_{\alpha},q} = \frac{|\boldsymbol{F}_{\nu_{\alpha},q}|}{cE_{\nu_{\alpha},q}} = \frac{|\boldsymbol{F}'_{\nu_{\alpha},q}|}{cE'_{\nu_{\alpha},q}} = {f}'_{\nu_{\alpha},q}
    \end{equation}
for $\nue$ and $\nuebar$ in the case of Eq.~(\ref{Eq:mom1}) and for $\nux$ in the case of Eq.~(\ref{Eq:mom2}).
Nevertheless, we have confirmed that our results are qualitatively unaffected by reasonable alternative prescriptions for the neutrino momentum behavior during flavor conversions.
A somewhat similar prescription was also recently used in Ref.~\cite{Just:2022flt} for studying the impact of FFCs during the evolution of neutron star merger remnants.

We finally remark that we switch on our schematic treatment of FFCs right at the beginning of the CCSN simulations, i.e., at the time when the stellar core collapse sets in.
However, applying FFCs during the collapse until core bounce does not have any relevant effect, because prior to bounce electron neutrinos $\nu_e$ dominate $\bar\nu_e$ and heavy-lepton neutrinos in number by several orders of magnitude, and therefore flavor conversions of the far subdominant $\nu_x$ cause negligibly small differences.
Of course, we cannot exclude that in a more realistic treatment, FFCs might find conditions to occur only at times considerably later than core bounce.
This adds additional uncertainty to our schematic treatment, in particular since we find the effects of FFCs to depend strongly on the phase of the post-bounce evolution.

\section{Simulation results}\label{sec:Results}

In this section, we present the results of a set of 1D simulations of our $20\,\Msol$ model.
We will refer to different simulations either as model with no flavor conversions (M-NFC), or we label them with the employed threshold density $\rho_{\mathrm{c}}$ for simulations where FFCs are applied in the region $\rho < \rho_{\mathrm{c}}$.
Since the values of $\rho_{\mathrm{c}}$ range between $10^9\,\gccm$ and $10^{14}\,\gccm$, the corresponding models are named M-1e09 to M-1e14.
Performing additional simulations with threshold densities of about $3\times10^9\,\gccm$, $3\times10^{10}\,\gccm$, $3\times10^{11}\,\gccm$, $3\times10^{12}\,\gccm$, and $3\times10^{13}\,\gccm$, we verified that our sampling is sufficiently dense, because these simulations do not reveal any qualitatively new insights.
Instead, their results match the systematic trends witnessed in the models presented in the following.

The discussion is divided into two parts.
The first part gives a comparative overview of the impact of flavor conversions on global model properties.
In the second part, we focus on more detailed aspects of some special simulations that are worth a closer look.

\subsection{Global Properties}
\label{sec:GlobalProperties}

Before reporting the findings of our simulations in detail, we will discuss the impact of neutrino FFCs on the energy transport in the CCSN environment in general terms.
We find that pair-wise flavor conversions modify the radiated and potentially detectable neutrino signal, change the energy loss of the stellar medium and the energy transfer between neutrinos and matter, and can thus alter the evolution of the SN shock and PNS.

\subsubsection{Overview of Main Effects}
\label{sec:MainEffects}

The effects of FFCs depend on the local strength of the weak-interaction coupling between neutrinos and matter and on whether \EtoX or \XtoE conversions occur.
Accordingly, we can discriminate between four different kinds of consequences.
All of them can play a role during the post-bounce evolution of the collapsing stellar core, because only after core bounce $\nuebar$ and $\nux$ are present in relevant abundances.
Moreover, the effects show up in different phases and in different spatial regions.

If flavor conversions take place only in the region ahead of the shock, they have no significant influence on the dynamics of CCSNe.
This can be understood by the fact that in the upstream region of the shock the densities are usually low and the infall velocities of the stellar gas are very high.
Under such conditions neutrino interactions with the stellar medium are infrequent and FFCs have no relevant feedback on the SN evolution.
In our set of simulations, model M-1e09 provides an example of such a case.

If flavor conversions take place in the cooling layer that extends from the neutrinospheric region to the gain radius,\footnote{The gain radius is located between the neutrinospheres and the stalled SN shock and is defined as the boundary between layers of net neutrino cooling below and net neutrino heating above.} they can enhance the cooling rate of the SN matter.
Conversions of the type \EtoX increase the abundance of $\nux$, which couple more weakly to the stellar medium and therefore lead to more efficient energy transport.
At the same time, the reduced $n_{\nue}$ and $n_{\nuebar}$ stimulate the production of $\nue$ and $\nuebar$ through $\beta$-processes in the settling accretion layer, enhancing neutrino cooling even further.
The accelerated energy loss through \EtoX conversions thus triggers a faster contraction of the PNS.

Despite this amplified cooling by FFCs, the inverse conversion of $\nux$ to $\nue$ and $\nuebar$ can result in the opposite effect, namely higher energy deposition in the heating layer between the gain radius and the shock.
This plays a role during phases when the spectral temperature of $\nux$ is higher than that of $\nue$ and $\nuebar$, in which case \XtoE conversion is favored for high neutrino energies.
The quadratic dependence of the charged-current absorption cross sections on the neutrino energy implies that a relatively smaller number of such conversions at high energies can potentially compensate the impact of \EtoX conversions at lower energies and can cause more heating in the gain layer.

Finally, potential flavor conversions deeper inside the PNS, interior to the neutrinospheres, can slow down the transport of energy there.
In the core of the PNS electron neutrinos are highly degenerate, for which reason the neutrino number densities obey the ordering $n_{\nue}>n_{\nux}>n_{\nuebar}$.
At such conditions \XtoE conversions can flatten the negative $n_{\nux}$ gradient that usually drives the cooling of the PNS by $\nux$ diffusion.
Consequently, neutrino conversions attenuate the energy loss from these regions because the $\nue$ and $\nuebar$ produced by $\nux$ conversions are more tightly coupled to the stellar plasma and quickly get absorbed by neutrons and protons.
If neutrino flavor conversions occur deep inside the PNS, the emission properties of the radiated neutrinos can be more strongly affected by changes of the PNS cooling than by the direct effects of flavor conversions.

When discussing the effects of neutrino flavor conversions in the SN interior, the factors of relevance are therefore the region where these conversions are assumed to take place and the spectral differences between $\nue$, $\nuebar$, and $\nux$, i.e., the relative differences in the number densities of these neutrino species at different energies.
The factors that influence the dynamical evolution of CCSNe, in particular of the SN shock, as a consequence of neutrino flavor conversions are the contraction behavior of the PNS and the neutrino heating in the gain layer downstream of the shock.

\begin{figure}
    \includegraphics{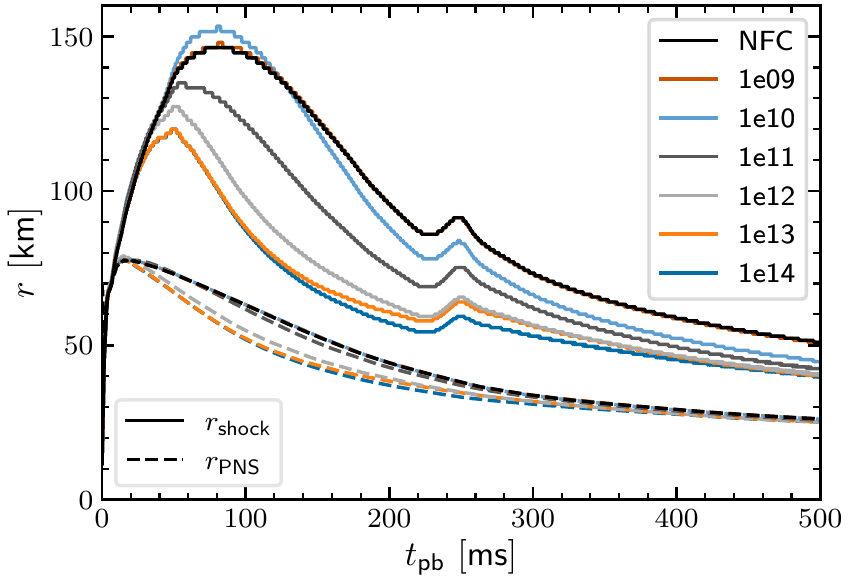}
    \caption{Shock radii (solid) and PNS radii (dashed) as functions of post-bounce time.
    The PNS radius is defined as the location where the density is $10^{11}\,\gccm$.
    The black curves represent the results of the simulation with no flavor conversions, model M-NFC, whereas the colored curves correspond to our set of models M-1e09 to M-1e14 with varied threshold densities for FFCs from $\rho_\mathrm{c} = 10^{9}\,\gccm$ to $\rho_\mathrm{c} = 10^{14}\,\gccm$.
    Note that the curves for the shock radii of models M-NFC and M-1e09 overlap.
    Also the curves for the PNS radii of models M-NFC, M-1e09, and M-1e10 lie on top of each other, and the PNS radii of models M-1e13 and M-1e14 are nearly identical.}
    \label{fig:ShockPNSRadius}
\end{figure}

\subsubsection{Evolution of Shock and Proto-Neutron Star Radii}
\label{sec:ShockPNSRadius}

The evolution of the shock radius with time, $r_{\mathrm{shock}}(t)$, in particular the maximum expansion of the shock, can serve as a rough indicator of the proximity of a CCSN simulation to explosion.
However, 1D results should be taken with great caution in this respect, because the absence of nonradial hydrodynamic instabilities in the postshock layer and convection inside the PNS can lead to qualitative differences of the dynamical behavior compared to multi-dimensional models.

Our simulations reveal a systematic and sensitive dependence of $r_{\mathrm{shock}}$ on the value of the threshold density $\rho_{\mathrm{c}}$.
The higher we choose this upper limit of the density regime where FFCs are assumed to take place, the weaker the shock expansion becomes, manifesting itself in a smaller maximum radius and a faster contraction of the shock after its maximum (Figure~\ref{fig:ShockPNSRadius}).
Basically, this behavior can be understood by a similar systematic ordering of the PNS radii, which are defined by the iso-density contours for a value of $\rho(r_\mathrm{PNS}) = 10^{11}\,\gccm$.
It is well known that in 1D simulations $r_{\mathrm{shock}}(t)$ and $r_{\mathrm{PNS}}(t)$ are tightly correlated because of the circumstance that the region between the PNS surface and the stalled shock possess a stratification that is nearly in hydrostatic equilibrium during the post-bounce accretion phase~\cite{Janka:2000bt,Marek:2007gr,Janka:2012wk}.

The shock and PNS radii of models M-NFC and M-1e09 evolve effectively identically.
This confirms our expectation that flavor conversions in the preshock region have a negligible influence on the CCSN evolution, which can be noticed also in all other dynamically relevant quantities.
In contrast, of course, the characteristic properties of the neutrino emission that can be measured by a distant observer reflect the effects of FFCs at densities $\rho < \rho_\mathrm{c} = 10^9\,\gccm$.

The shock in models M-1e11 to M-1e14 contracts much faster because in these simulations flavor conversions \EtoX happen in the near-surface layers of the PNS.
As discussed before, $\nux$ are less tightly coupled to the stellar medium than $\nue$ and $\nuebar$ and therefore the FFCs enhance the neutrino cooling of these layers.
Because of the enhanced loss of internal energy in the cooling region, the gravitational settling of the PNS accretion layer is accelerated, the PNS radius contracts more rapidly, and the radius of the accretion shock follows accordingly.

\begin{figure}
    \includegraphics{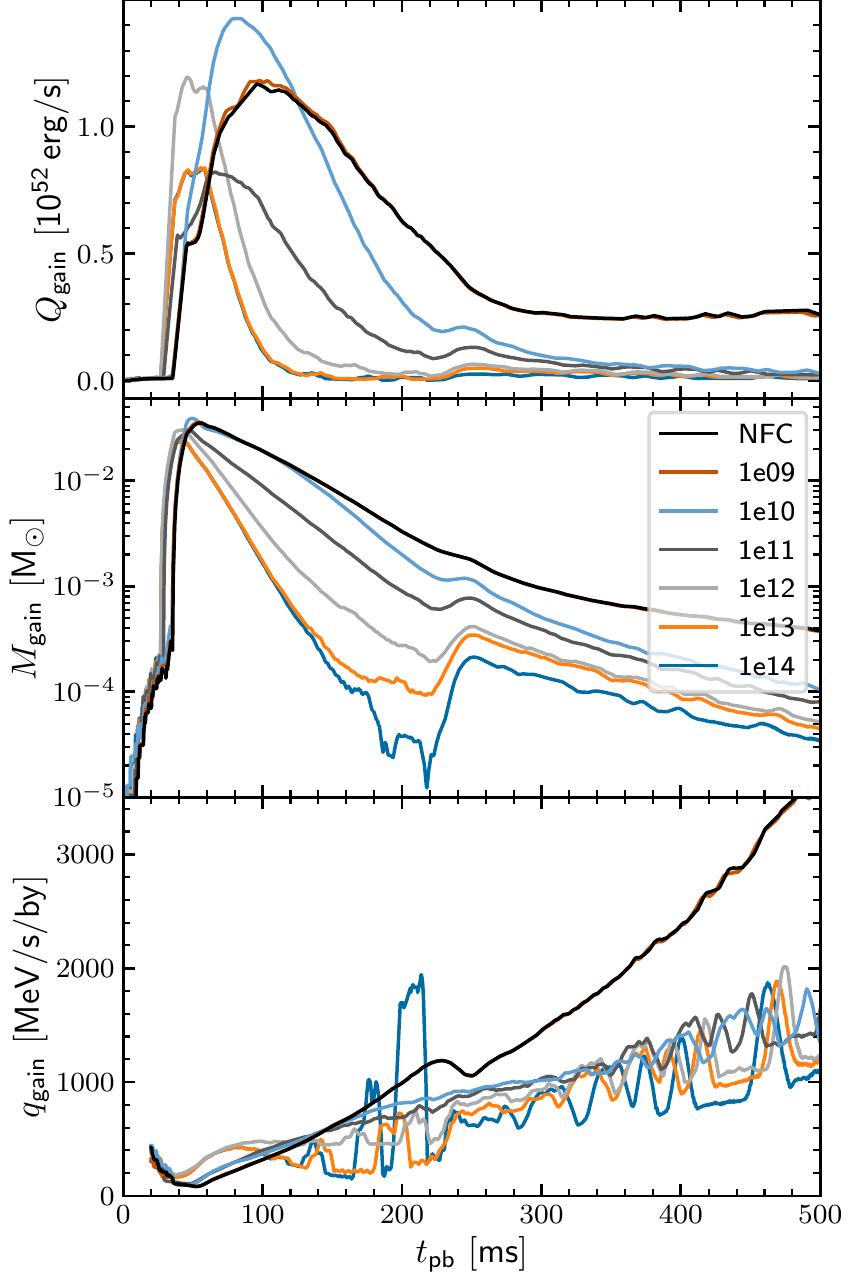}
    \caption{Total neutrino heating rate, mass, and mass-specific heating rate in the gain layer (from top to bottom) as functions of post-bounce time.
    The color coding is the same as in Figure~\ref{fig:ShockPNSRadius} and short-time fluctuations are reduced by smoothing the curves with a running average of 10\,ms.
    The high excursions of $q_{\mathrm{gain}}$ in model M-1e14 between $t_\mathrm{pb}\sim 170$\,ms and $\sim$220\,ms are caused by the narrowness of the gain layer due to the small shock radius in this time interval, which leads to an ill-determined gain-layer mass.
    (Model M-1e13 in Figure~\ref{fig:massshell} possesses a similarly narrow gain layer.)
    Note that the curves of models M-NFC and M-1e09 mostly coincide.}
    \label{fig:Heating_gain}
\end{figure}

Model M-1e10 exhibits a slightly different shock evolution.
In the early post-bounce phase ($t_\mathrm{pb}$ between $\sim$40\,ms and $\sim$130\,ms) the shock expands to a slightly larger radius than in model M-NFC and shows a faster contraction only at later times.
The reason for this difference is connected to the fact that initially $r(\rho_{\mathrm{c}})$ is located in the gain layer.
Therefore flavor conversions do not lead to enhanced cooling of the near-surface layers of the PNS and, correspondingly, the PNS radius of model M-1e10 tracks the behavior of the PNS radius in model M-NFC.
Instead, during the early post-bounce phase increased neutrino heating in the gain layer (Figure~\ref{fig:Heating_gain}) supports the stronger shock expansion.
The increased heating can be explained by \XtoE conversions of high-energy $\nux$ in the gain layer.
Since the original $\nux$ spectra of model M-NFC are harder than those of $\nue$ and $\nuebar$ (i.e., their mean energies are higher; Figure~\ref{fig:Mean_energy}), the conversion leads to higher $\nue$ and $\nuebar$ energies and therefore more energy deposition in the gain layer, despite a reduction of the $\nue$ and $\nuebar$ luminosities (Figure~\ref{fig:Luminosity}).

The differences in the gain-layer heating and the neutrino properties between the simulations of our model set will be further discussed in Sections~\ref{sec:Heating} and \ref{sec:NeutrinoEmissionProperties}, respectively, and more details on the evolution of model M-1e10 will be given in Section~\ref{sec:1e10}.

\subsubsection{Neutrino Heating in the Gain Region}
\label{sec:Heating}

In order to better judge the effects of FFCs on the neutrino heating in the gain region, we consider the time evolution of three quantities: the total heating rate in the gain layer, $Q_{\mathrm{gain}}$, the mass in the gain layer, $M_{\mathrm{gain}}$, and the specific heating rate, i.e., $q_\mathrm{gain} = Q_\mathrm{gain}/M_\mathrm{gain}$, see Figure~\ref{fig:Heating_gain}.

Indeed, the FFCs lead to major changes of $Q_\mathrm{gain}$ (top panel), but these differences can be attributed to a large extent to differences in $M_\mathrm{gain}$ (middle panel).
These are a consequence of the shrinking shock and gain radii, which adjust to the faster PNS contraction.
The specific heating rate $q_\mathrm{gain}$ more directly reflects the impact of FFCs on the relevant neutrino properties that determine the neutrino energy deposition in the postshock layer.
For example, $q_\mathrm{gain}$ is increased for up to about 150\,ms after bounce in the simulations with flavor conversion compared to model M-NFC, despite the fact that the total heating rate is reduced during a part of this time interval.

\begin{figure}
    \includegraphics{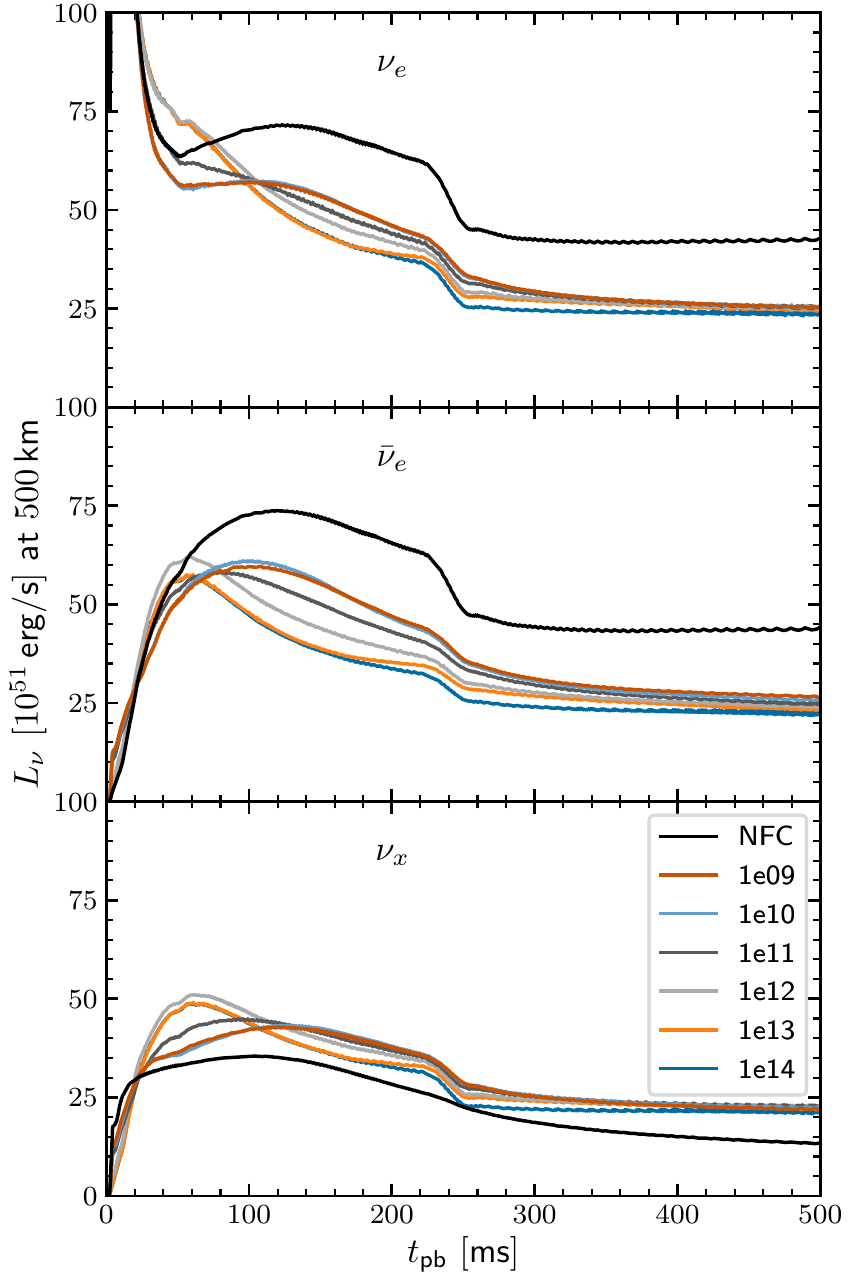}
    \caption{Neutrino luminosities as functions of post-bounce time.
    They are measured at a radius of 500\,km and transformed to an observer in the lab frame at infinity.
    The color coding is the same as in Figure~\ref{fig:ShockPNSRadius}.
    The top, middle, and bottom panels show $L_{\nue}$, $L_{\nuebar}$, and $L_{\nux}$, respectively.
    $L_{\nux}$ is the luminosity for one species of heavy-lepton neutrinos.
    In the top panel the maximum of the $\nue$ burst has been cut off for better visibility.
    It peaks at about $5.5\times10^{53}$\,erg\,s$^{-1}$ with a variation of $\sim$10\% of the maximum value between the different models.
    Note that the lines for models M-1e09 and M-1e10 on the one hand and for models M-1e13 and M-1e14 on the other hand partly overlap.
    The step-like decline of the luminosities between about 230\,ms and 250\,ms is caused by the fast decrease of the mass accretion rate onto the PNS when the Si/O interface falls through the stalled shock.
    This leads to a corresponding drop of the accretion luminosities of $\nue$ and $\nuebar$.}
    \label{fig:Luminosity}
\end{figure}

\begin{figure}
    \includegraphics{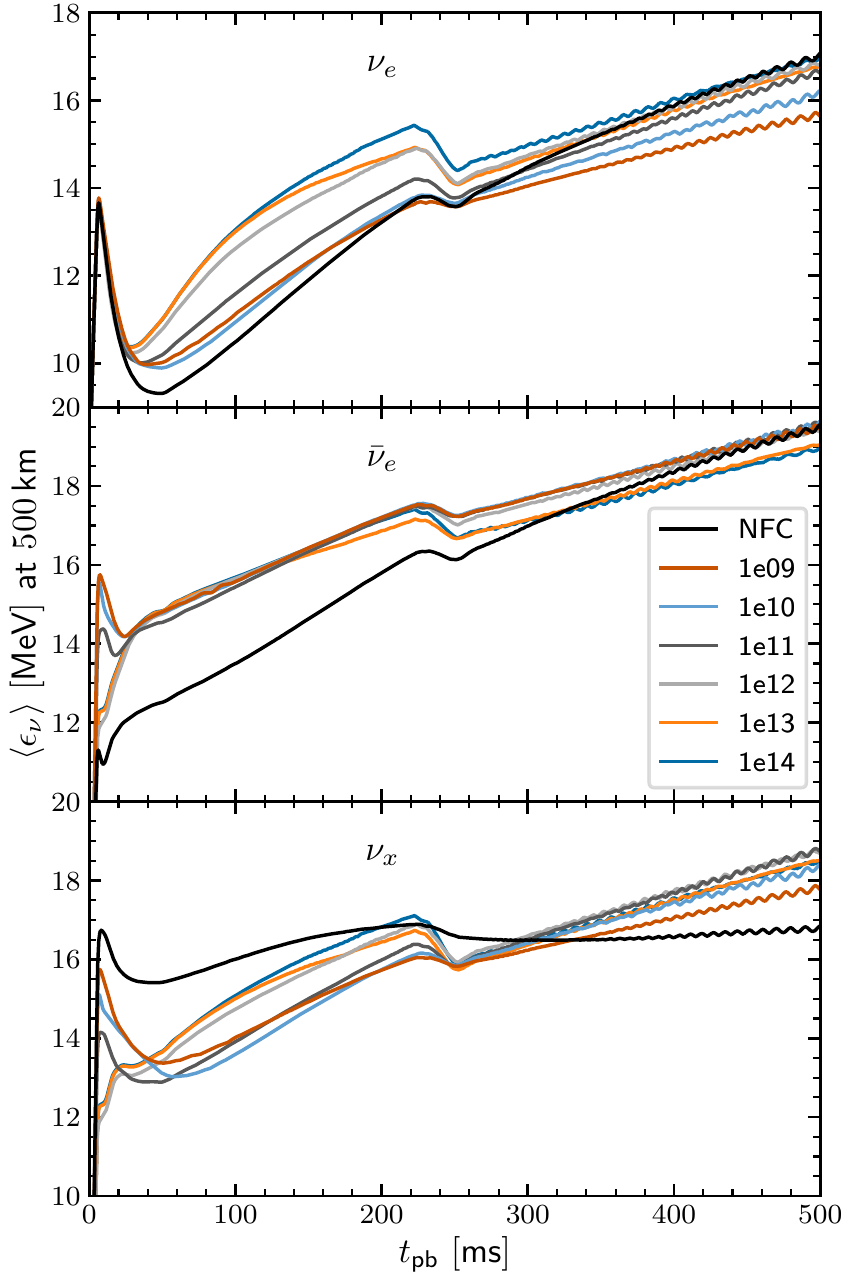}
    \caption{Mean energies of the radiated neutrino spectra versus post-bounce time, analogous to the neutrino luminosities of Figure~\ref{fig:Luminosity}.
    The brief, non-monotonic behavior between about 230\,ms and 250\,ms is again connected to the decline of mass accretion rate when the infalling Si/O interface passes the stalled shock.
    The curves have been smoothed with a running average of 5\,ms to improve readability.}
    \label{fig:Mean_energy}
\end{figure}

There are two reasons for the relative enhancement of $q_\mathrm{gain}$.
First, it can be explained by \XtoE conversions occurring at higher neutrino energies in the early post-bounce phase then the mean energies of $\nux$ are bigger than those of $\nue$ and $\nuebar$.
The corresponding hardening of the $\nue$ and $\nuebar$ spectra in models with FFCs increases the postshock heating, despite reduced $\nue$ and $\nuebar$ luminosities (see Figures~\ref{fig:Luminosity} and \ref{fig:Mean_energy} for the neutrino quantities).
This is particularly relevant in models M-1e10 and M-1e11.
However, with respect to the shock dynamics, the stronger heating may be compensated by enhanced cooling in the cooling layer interior to the gain radius, for which reason the shock radius in model M-1e11 is smaller than in models M-NFC and M-1e10.
The second reason for higher specific heating rates $q_\mathrm{gain}$ during the earliest post-bounce evolution ($t_\mathrm{pb} \lesssim 100$\,ms) is connected to the accelerated energy loss through $\EtoX$ conversions in the neutrinospheric region of the PNS.
This triggers a much more rapid PNS contraction in models M-1e12, M-1e13, and M-1e14 compared to all other models (see Figure~\ref{fig:ShockPNSRadius}) and thus, indirectly, also higher $\nue$ and $\nuebar$ luminosities in addition to a most extreme boosting of the $\nux$ luminosities (Figure~\ref{fig:Luminosity}) and in addition to the significantly increased mean neutrino energies, in particular of $\nue$ (Figure~\ref{fig:Mean_energy}).
The correspondingly strong amplification of $q_\mathrm{gain}$ in models M-1e12 to M-1e14 can be witnessed in Figure~\ref{fig:Heating_gain} for 100--120\,ms, but it has no noticeable supportive effect on the shock because of the extraordinarily rapid contraction of the PNS.

At times $t_\mathrm{pb}\gtrsim 150$\,ms, at latest, not only the total but also the specific neutrino heating rates in the gain layer of all models with FFCs are lower than in model M-NFC.
This is mainly connected to the fact that the differences between the mean energies $\langle\epsilon_\nu\rangle$ of $\nue$, $\nuebar$, and $\nux$ decrease with progressing post-bounce time in model M-NFC (Figure~\ref{fig:Mean_energy}), which diminishes the impact of \XtoE conversions on the gain-layer heating.

\subsubsection{Neutrino Luminosities and Mean Energies}
\label{sec:NeutrinoEmissionProperties}

Figures~\ref{fig:Luminosity} and \ref{fig:Mean_energy} display the luminosities and mean energies of the radiated neutrinos as functions of post-bounce time, measured at a radius of 500\,km and properly transformed to an observer in the lab frame at infinity.

Two models serve us as reference cases for our comparative discussion.
On the one hand, there is model M-NFC for the results without flavor conversions, and on the other hand there is model M-1e09, where flavor conversions occur in the preshock region and change the neutrino emission properties, but these changes neither have any feedback on the CCSN dynamics nor are they affected by variations in the SN evolution.

The comparison of the luminosities clearly shows that the dominant conversion channel is \EtoX.
The luminosities in models M-1e10 and M-1e09 are almost identical, which means that the enhanced gain-layer heating and slightly different shock evolution in the former have little effect on the observable neutrino fluxes.
In contrast, the enhanced luminosities mainly of $\nue$ and $\nux$ and more moderately also of $\nuebar$ in models M-1e12 to M-1e14 during the first $\sim$100\,ms after bounce are a consequence of the faster PNS and shock contraction.
Due to the accelerated settling of the accretion mantle of the PNS, more gravitational binding energy is carried away by neutrinos.
Since the faster PNS contraction goes hand in hand with a faster deleptonization, $L_{\nue}$ can even exceed the $\nue$ luminosity in model M-NFC for a brief period of time, despite the \EtoX conversions.
In model M-1e11 similar trends, though much less extreme, can be witnessed.
At times later than $t_\mathrm{pb} \sim 120$\,ms, when the PNS contraction becomes weaker, the luminosities of all neutrino species get ordered with the value of $\rho_\mathrm{c}$ such that lower luminosities correlate with higher values of the threshold density for FFCs.
Because of the dominant \EtoX conversion, $L_{\nue}$ and $L_{\nuebar}$ are reduced compared to those of model M-NFC in all cases and at all later post-bounce times, whereas $L_{\nux}$ is higher in all models with flavor conversions.

Despite this reduction of the $\nue$ and $\nuebar$ luminosities and increase of the $\nux$ luminosity connected with FFCs, the mean neutrino energies, $\langle\epsilon_\nu\rangle$, in Figure~\ref{fig:Mean_energy} display the opposite effects until 200--300\,ms after bounce (the exact time depends on the neutrino species), i.e., $\langle\epsilon_{\nue}\rangle$ and $\langle\epsilon_{\nuebar}\rangle$ are increased, whereas $\langle\epsilon_{\nux}\rangle$ is decreased in all models with FFCs compared to model M-NFC.
This is a consequence of mainly \EtoX conversions in the bulk of the spectra and additional \XtoE conversions at high neutrino energies.

However, at times later than 200--300\,ms after bounce the effect is reversed, i.e., the mean energies of $\nue$ and $\nuebar$ in models with FFCs begin to drop below those of model M-NFC, whereas for $\nux$ the opposite situation occurs.
This can be understood by the fact that between 200\,ms and 400\,ms after bounce the ordering of the mean neutrino energies in model M-NFC changes from initially
$\langle\epsilon_{\nux}\rangle>\langle\epsilon_{\nuebar}\rangle>\langle\epsilon_{\nue}\rangle$
first to
$\langle\epsilon_{\nuebar}\rangle>\langle\epsilon_{\nux}\rangle>\langle\epsilon_{\nue}\rangle$
and afterwards to
$\langle\epsilon_{\nuebar}\rangle>\langle\epsilon_{\nue}\rangle>\langle\epsilon_{\nux}\rangle$.
The continuous growth of the mean energies of $\nue$ and $\nuebar$ is explained by the ongoing accretion of infalling gas onto the PNS.
The corresponding conversion of gravitational binding energy to internal energy leads to a monotonic growth the temperature in the neutrinospheric region of $\nue$ and $\nuebar$, whereas the temperature at the deeper neutrinosphere of $\nux$ remains nearly constant.

It is interesting that after an initial, more complex phase of $\sim$40\,ms, the mean energies in particular of $\nue$ and $\nux$ reveal a similar systematic ordering with $\rho_\mathrm{c}$ as we found for the neutrino luminosities, however now in the opposite direction and with some minor deviations.
The higher the value of $\rho_\mathrm{c}$ is, the higher the values of $\langle\epsilon_{\nue}\rangle$ and $\langle\epsilon_{\nux}\rangle$ are.
The effect is particularly pronounced for models M-1e12, M-1e13, and M1-e14, where the PNS contraction is fastest and has the strongest impact on the properties of the neutrino emission.
In contrast to the luminosities, the trend of the mean energies includes model M-1e09; only model M-1e10 partly breaks the systematics, and models M-1e12 and M-1e13 slightly do so at $t_\mathrm{pb}\gtrsim 180$\,ms, and in the case of $\langle\epsilon_{\nux}\rangle$ also model M-1e11 shows deviations until about 130\,ms and at $t_\mathrm{pb}\gtrsim 280$\,ms.
The reason for the energy ordering is again connected to the strength of the PNS contraction and its influence on the temperatures at the neutrinospheres (for models M-1e12--M1e14), combined with the dominant \EtoX conversions in the bulk of the $\nue$ and $\nux$ spectra around their spectral peaks on the one hand and the \XtoE conversions in the high-energy spectral tails on the other hand.
These conversions lead to an increase of $\langle\epsilon_{\nue}\rangle$ with higher threshold densities and similarly for $\langle\epsilon_{\nux}\rangle$, though the latter mean energies are always below those of model M-NFC until about 300\,ms after bounce.

\begin{figure}
    \includegraphics{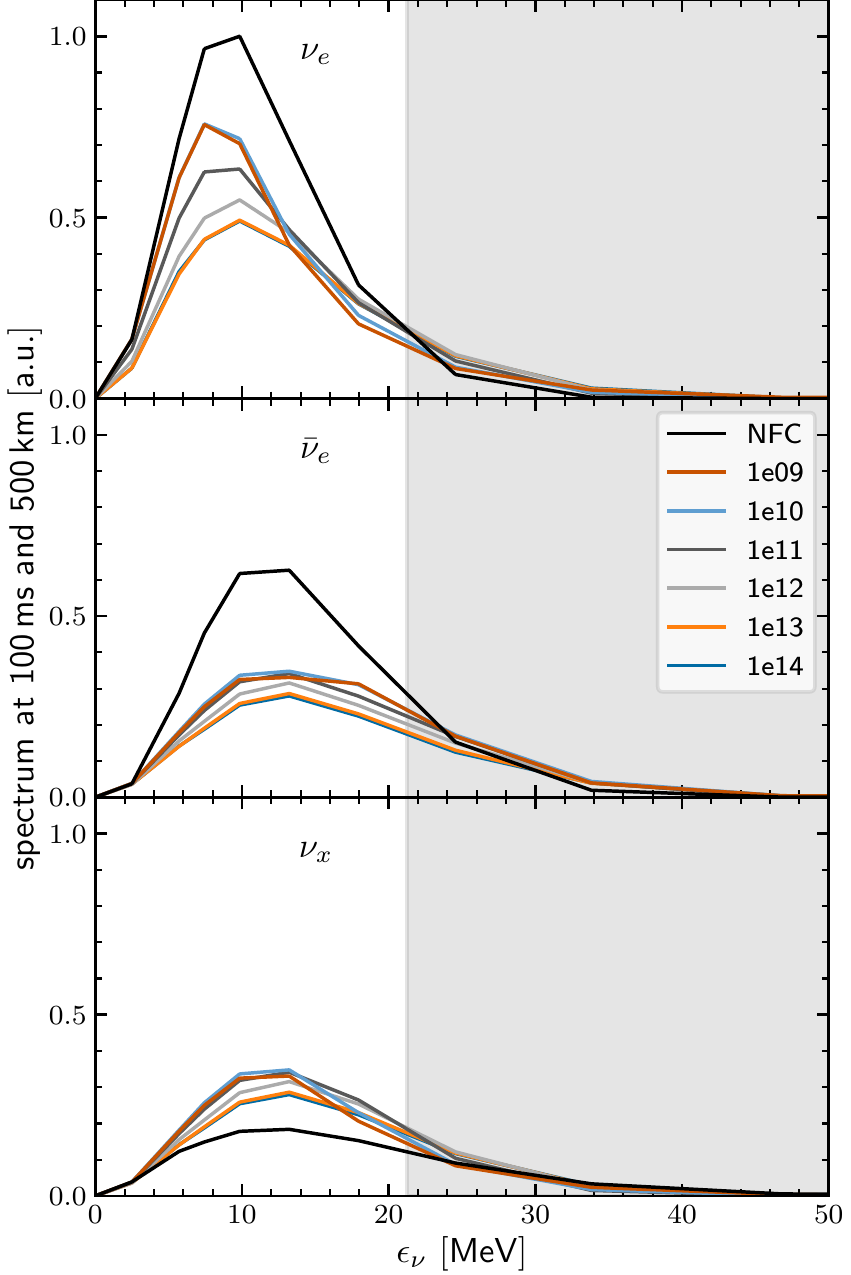}
    \caption{Neutrino spectra at 500\,km and $t_\mathrm{pb}=100$\,ms, normalized by the maximum peak value of the $\nue$ spectrum obtained in all simulations, i.e. model M-NFC.
    The color coding is the same as in Figure~\ref{fig:ShockPNSRadius}.
    The type of flavor conversions depends on $\epsilon_\nu$.
    It is dominantly \XtoE conversion in the high-energy tails of the spectra, roughly indicated by the gray shading, and mainly \EtoX conversion in the unshaded region.
    Note that the curves partially overlap.
    This is the case especially for models M-1e09 and M-1e10 and also models M-1e13 and M-1e14.}
    \label{fig:Spectrum}
\end{figure}

The neutrino spectra of the radiated neutrinos at a distance of $r=500\,\unit{km}$ and post-bounce time of $t_{\mathrm{pb}}=100\,\unit{ms}$, normalized by a common factor, are displayed in Figure~\ref{fig:Spectrum}.
It can be seen that in the high-energy spectral tails, roughly indicated by gray shading, the dominant conversion is \XtoE, whereas it is \EtoX in the main parts of the spectra at lower neutrino energies, as mentioned earlier.
It is an interesting question, deserving further studies, whether the lifting of the high-energy tails of the $\nue$ and $\nuebar$ spectra could provide an identifying observational signature of FFCs, in particular because of the boosting of the high-energy tails of the $\nux$ spectra ($\epsilon_{\nux} \gtrsim 50$\,MeV) through neutrino shock acceleration in failed CCSNe \cite{Nagakura:2020gls}, and potentially despite the flavor mixing associated with matter resonances and vacuum propagation.

\begin{figure*}
    \includegraphics[width=\textwidth]{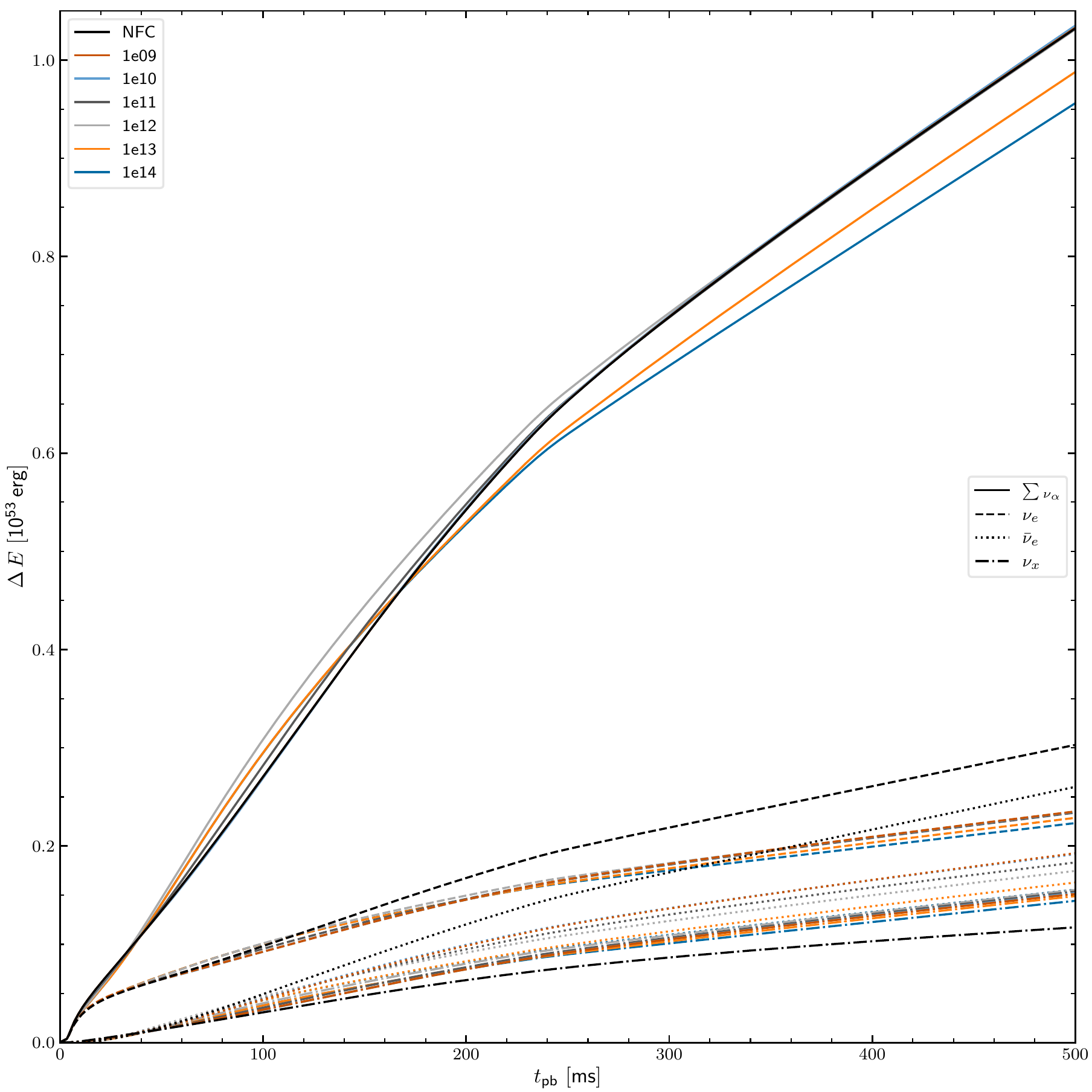}%
    \vskip-6pt
    \caption{Time integrated energy loss due to neutrino emission, $\Delta E$, evaluated with the luminosities measured at 500\,km and transformed to a lab-frame observer at infinity.
    The color coding is the same as in Figure~\ref{fig:ShockPNSRadius}.
    The solid lines represent the total energy loss, summed up for neutrinos and antineutrinos of all flavors.
    The dashed, dotted, and dash-dotted lines correspond to the individual energies lost in $\nue$, $\nuebar$, and one species of heavy-lepton neutrinos, $\nux$, respectively.
    Note that the lines of various models lie on top of each other and cannot be discriminated.}
    \label{fig:Energy_loss}
\end{figure*}

Many of the effects that we have described so far also leave imprints on the time-integrated energies lost by the PNS through the emission of neutrinos of different species, $\Delta E_\nu(t_\mathrm{pb}) = \int_{0}^{t_{\rm{pb}}} \mathrm{d}t\ L_{\nu}(t)$, and on the sum of these energies, $\Delta E(t_\mathrm{pb}) = \Delta E_{\nue} + \Delta E_{\nuebar} + 4\Delta E_{\nux}$ (Figure~\ref{fig:Energy_loss}).

For all models with FFCs, $\Delta E_{\nue}$ and $\Delta E_{\nuebar}$ are significantly lower and $\Delta E_{\nux}$ is significantly higher than in model M-NFC, as expected from the dominant \EtoX conversion in the bulk of the spectra during most of the evolution.
But there are exceptions during the early post-bounce phase, $t_\mathrm{pb} \lesssim 100$\,ms, which are connected to several effects.
At very early times, $t_\mathrm{pb} \lesssim 20$\,ms, \XtoE is the leading flavor conversion process because initially the production rate of $\nux$ is higher than that of $\nuebar$ due to the high $\nue$ degeneracy in the post-bounce deleptonization phase (see Figure~\ref{fig:Luminosity} for the impact of FFCs on the $\nuebar$ and $\nux$ luminosities during this phase).
In addition, as mentioned before, the extremely fast contraction of the PNS in models M-1e12, M-1e13, and M-1e14 boosts all luminosities including those of $\nue$ and $\nuebar$ above the values of model M-NFC for a time interval up to $\sim$80\,ms after bounce (Figure~\ref{fig:Luminosity}).

Concerning the total energy release in all neutrino species, $\Delta E(t_\mathrm{pb})$, models M-NFC, M-1e09, and M-1e10 are effectively identical at all times, because the simulations show nearly the same dynamical evolution and small differences connected to neutrino absorption in the gain layer have little influence on the total energy lost in the neutrino emission.

In models M-1e11 and M-1e12, $\Delta E(t_\mathrm{pb})$ is initially (up to $t_\mathrm{pb}\simeq 250$\,ms and $t_\mathrm{pb}\simeq 325$\,ms, respectively) higher, and then approaches the value of model M-NFC.
For these models, we have argued that the flavor conversions lead to enhanced cooling of the outer layers of the PNS because of an increased rate of energy transport as a consequence of \EtoX conversions.
The corresponding early increase of $\Delta E_{\nux}$ accelerates the settling of the near-surface layers of the PNS and steepens the growth of $\Delta E(t_\mathrm{pb})$ in general.
Once the settling of the PNS's accretion mantle in models M-NFC, M-1e09, and M-1e10 catches up, $\Delta E(t_\mathrm{pb})$ of the five models becomes more and more equal and the subsequent evolution proceeds nearly identically.

Similarly, the time-integrated total neutrino energy release of models M-1e13 and M-1e14 is also higher than in model M-NFC up to $t_\mathrm{pb} \simeq 180$\,ms, but this excess in the energy loss compared to model M-NFC is reversed at later times.
In fact, the curves for $\Delta E(t_\mathrm{pb})$ of models M-1e13 and M-1e14 nearly coincide for a somewhat longer time interval and differences between both of them become clearly visible in Figure~\ref{fig:Energy_loss} only at $t_\mathrm{pb}\gtrsim 230$\,ms.
In these models flavor conversions occur not only in the outer accretion layer, where \EtoX conversions speed up the energy loss, but also in deeper regions of the PNS, where \XtoE conversions are favored because of the high $\nue$ degeneracy (see Section~\ref{sec:MainEffects}).
In these regions inside the PNS, at densities above a few $10^{12}$\,g\,cm$^{-3}$, $\nux$ play an important role for the energy transport and the net effect of \XtoE conversions is a deceleration of the energy transport out of the high-density core of the PNS.
This reduces $\Delta E(t_\mathrm{pb})$ in models M-1e13 and M-1e14 compared to all other simulations at $t_\mathrm{pb}\gtrsim 180$\,ms.
The effect becomes prominent only at such late times, because the neutrino diffusion out of the core is relatively slow, and therefore the release of gravitational binding energy by neutrinos produced in the settling accretion layer is more important in the earlier post-bounce phase.
As time progresses, the neutrino energy loss of model M-1e14 falls more and more behind that of model M-1e13, compatible with the fact that the energy transport is decelerated in a larger region in the former simulation.
However, in the long run, $\Delta E$ of both models will approach the value of the other models, because the final value of the energy release depends only on the NS mass and the nuclear EoS.

Besides changing the neutrino emission properties, the neutrino cooling and heating, and the PNS and shock evolution, FFCs also have an impact on the electron fraction $Y_e$ in the different regions behind the shock.
While these $Y_e$ changes are transient effects in matter that gets accreted into the final NS, where ultimately neutrino-less beta-equilibrium will be established, it may be more interesting to consider possible consequences of FFCs on the electron fraction in neutrino-heated ejecta in the case of successful CCSN explosions.
Naturally, this question can be reliably addressed only on grounds of simulations that produce explosions, because the onset of the CCSN blast also alters the neutrino emission of the PNS.
For a tentative assessment, potentially indicating trends, we evaluate Equation~(10) of Ref.~\citep{Janka+2023}, using the neutrino luminosities and mean energies at a radius of 500\,km and at $t_\mathrm{pb}=500\,\mathrm{ms}$ after bounce.
Thus we estimate an asymptotic value of $Y_e = 0.574$ for a neutrino-driven mass outflow in model M-NFC, whereas $Y_e$ ranges between $0.552$ and $0.599$ for outflows in our models with FFCs, monotonically increasing from M-1e09 with the lowest value to M-1e14 with the highest value.
This suggests that FFCs might change $Y_e$ within a few percent of the NFC value in both directions but not enough to convert proton-rich ejecta of model M-NFC to neutron rich ones in our investigated cases.

\subsection{Additional Aspects of Individual Simulations}

In this section we take a closer look at some special features of individual simulations that deserve a more detailed inspection.
In the first part we will discuss the effects of flavor conversions happening in the heating and cooling layers around the gain radius in model M-1e10.
In the second part, we will illuminate the consequences how the deceleration of the neutrino energy transport by \EtoX conversions in the dense regions well inside the neutrinospheres affects the PNS structure in models M-1e13 and M-1e14.

\begin{figure}[t!]
    \includegraphics{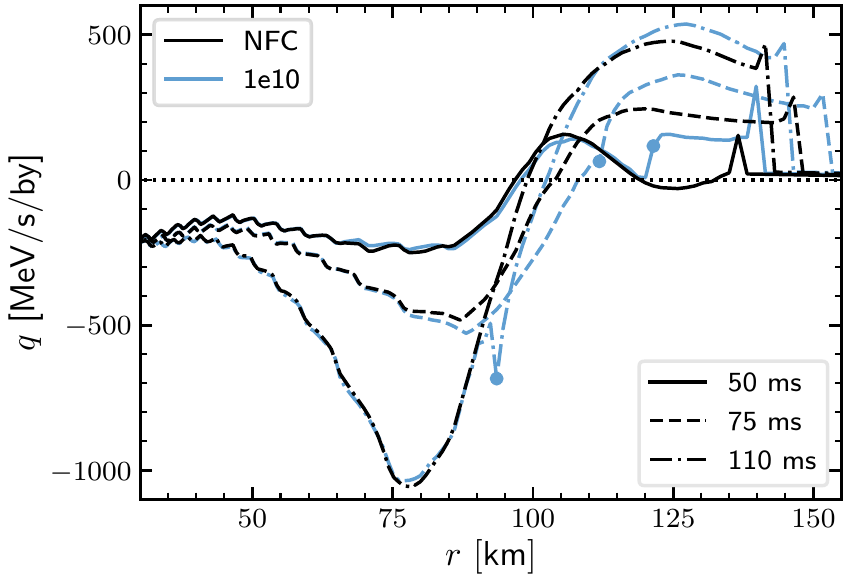}
    \caption{Radial profiles of the specific neutrino heating and cooling rate for model M-1e10 compared to model M-NFC at different post-bounce times.
    The color coding is the same as in Figure~\ref{fig:ShockPNSRadius}.
    For model M-1e10 the bullets mark the radial locations where $\rho_\mathrm{c} < 10^{10}$\,g\,cm$^{-3}$, i.e., they mark the first computational zone where FFCs are assumed to take place.
    The corresponding radii are 120\,km at $t_\mathrm{pb} = 50$\,ms, 112\,km at $t_\mathrm{pb} = 75$\,ms, and 93\,km at $t_\mathrm{pb} = 110$\,ms.}
    \label{fig:Heating_rate_profile}
\end{figure}

\subsubsection{FFCs in the Cooling and Heating Regions of Model M-1e10}
\label{sec:1e10}

Model M-1e10 provides more detailed insights into the effects connected to flavor conversions in the cooling and heating regions between PNS and stalled SN shock.
As described in Section~\ref{sec:ShockPNSRadius}, this simulation is special because during the initial phase of shock expansion, the shock radius $r_\mathrm{shock}$ reaches a larger maximum value than in model M-NFC.
But this greater extension cannot be maintained and eventually, $r_{\mathrm{shock}}$ retreats even faster.
In this model the radius exterior to which FFCs are assumed to occur is $r(\rho_\mathrm{c}) \gtrsim r_\mathrm{shock}$ until $t_\mathrm{pb} \simeq 35\,\unit{ms}$.
At that time $r(\rho_\mathrm{c})$ drops below $r_\mathrm{shock}$, and shortly afterwards ($t_\mathrm{pb} \simeq 39$\,ms) a gain layer forms behind the stalled SN shock.
As time passes, $r(\rho_\mathrm{c})$ moves even farther inward and retreats into the cooling region below the gain radius at $t_\mathrm{pb} \simeq 78$\,ms.

Figure~\ref{fig:Heating_rate_profile} visualizes this evolution and its impact on the local specific neutrino heating and cooling rate, $q(r)$, for model M-1e10 in comparison to model M-NFC at times before, during, and after $r(\rho_\mathrm{c})$ shifts into the cooling region.
Although the matter profiles of the NFC and 1e10 models are slightly shifted due to the different positions of $r_\mathrm{shock}$, one can see that flavor conversions tend to increase the heating exterior to the gain radius and the cooling below the gain radius.

As long as $r(\rho_\mathrm{c}) > r_\mathrm{shock}$, there is no relevant influence of FFCs on the CCSN dynamics, in agreement with what we discussed for model M-1e09.
When $r(\rho_\mathrm{c})$ moves into the neutrino heating layer (solid and dashed lines in Figure~\ref{fig:Heating_rate_profile}), \XtoE conversions of high-energy $\nux$ lead to enhanced absorption of $\nue$ and $\nuebar$ via the $\beta$-reactions.
This increases the specific net heating rate $q$ in the gain layer despite the overall reduction of the $\nue$ and $\nuebar$ luminosities by \EtoX conversions in the bulk of the energy spectra and the corresponding rise of the $\nux$ luminosities (see Sections~\ref{sec:Heating} and \ref{sec:NeutrinoEmissionProperties}).
When $r(\rho_\mathrm{c})$ shifts inward below the gain radius, the \EtoX conversions of the lower-energy neutrinos reduce the $\nue$ and $\nuebar$ abundances in the cooling layer and permit additional production of these neutrinos by electron and positron captures because of sufficiently high temperatures in that region.
Therefore the specific net cooling rate $q$ is enhanced in the cooling layer (dash-dotted line in Figure~\ref{fig:Heating_rate_profile}), which triggers the faster shock contraction after the maximum shock expansion in model M-1e10.

The described effects may explain the difference between our results and previous studies performed to explore the impact of neutrino flavor conversions on the CCSN physics, e.g., in Refs.~\cite{Stapleford:2019yqg, Dasgupta:2011jf}.
While these earlier works considered slow modes of neutrino flavor conversion in the first place, they also allowed for flavor conversions in a much more limited region of the SN core, constrained to the zones very close to the shock or exterior to it.
Hence, these studies observed a relatively weak impact of flavor conversions on the SN evolution.

\begin{figure}
    \includegraphics{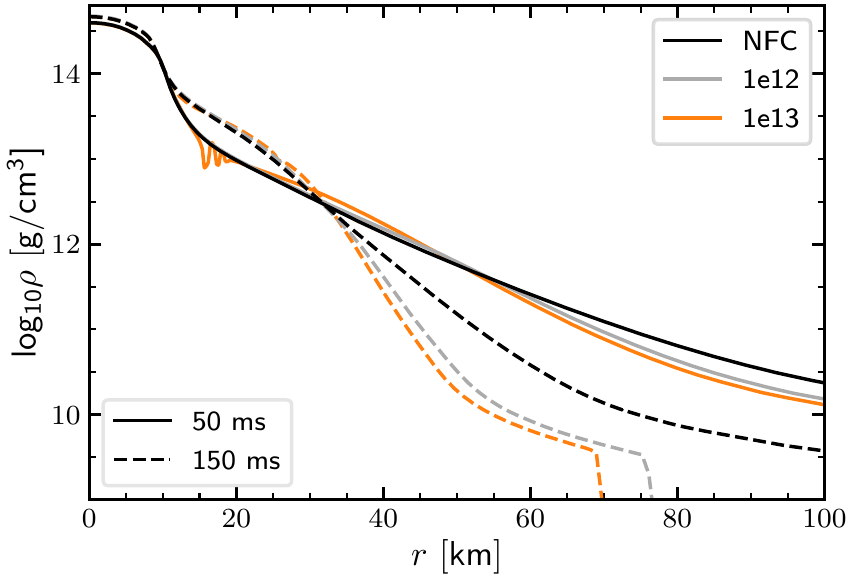}\\
    \includegraphics{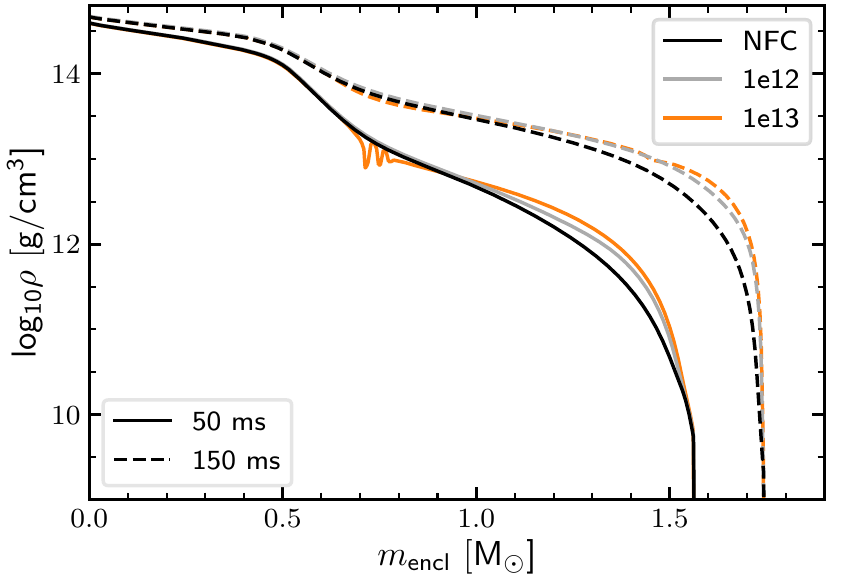}
    \caption{Profiles of the mass density in models M-NFC, M-1e12, and M-1e13 versus radius $r$ (top) and enclosed mass $m_\mathrm{encl}$ (bottom) at 50\,ms and 150\,ms after core bounce.
    The color coding is the same as in Figure~\ref{fig:ShockPNSRadius}.
    In both simulations of M-1e12 and M-1e13, the contraction of the outer layers of the PNS is faster than in model M-NFC, with more extreme effects in model M-1e13 (top panel).
    Therefore, FFCs lead to higher densities and more mass in the mantle region of the PNS and a steeper density decline above this layer.
    It should also be noticed that their absence at $\rho > 10^{12}$\,g\,cm$^{-3}$ in model M-1e12 permits the formation of a more compact high-density core ($m_\mathrm{encl} \lesssim 1.2$\,M$_\odot$ and $r\lesssim 15$\,km) earlier than in model M-1e13, where the contraction of the inner region is delayed because of the higher reservoir of internal energy.
    }
    \label{fig:Den_profile}
\end{figure}

\subsubsection{FFCs at High PNS Densities in Models M-1e13 and M-1e14}
\label{sec:HighRho}

Models M-1e13 and M-1e14 provide a natural extension of our parametric study for cases where FFCs also occur in the high-density interior ($\rho_\mathrm{c}\gtrsim 10^{13}$\,g\,cm$^{-3}$) of the PNS, well inside the neutrinospheres.
This situation is quite interesting because it leads to effects that are qualitatively different from those witnessed for lower values of $\rho_\mathrm{c}$.

At densities below $\sim$$10^{12}$\,g\,cm$^{-3}$, \EtoX are the dominant flavor conversions and lead to accelerated energy loss from the accretion layer of the PNS because of the looser coupling between $\nux$ and the stellar medium.
Consequently, the PNS radius exhibits a faster contraction (Figure~\ref{fig:ShockPNSRadius}).
Instead, between $10^{12}$\,g\,cm$^{-3}$ and $10^{13}$\,g\,cm$^{-3}$ \XtoE conversions become the dominant process because of the increasing $\nue$ degeneracy and the corresponding underabundance of $\nuebar$.
Since $\nu_x$ play an important role for the energy transport out of such high-density regions, the \XtoE conversions imply a significant reduction of the energy transport deep inside the PNS.
The thus decelerated loss of energy from the PNS core can explain the evolution of $\Delta E(t)$ in models M-1e13 and M-1e14 seen in Figure~\ref{fig:Energy_loss}.
Initially, the more efficient energy loss from the outer layers of the SN core leads to a steeper increase of $\Delta E(t)$ compared to model M-NFC, and later, when the neutrino diffusion out of the core becomes more important, the retarded transport of energy deep inside the PNS causes a slower growth of $\Delta E(t)$.

Interestingly, the decelerated loss of energy also modifies the radial structure in the high-density interior of the PNS.
This is visible in Figure~\ref{fig:Den_profile}, where we plot density profiles vs.\ radius and enclosed mass of model M-1e13 compared to those of models M-1e12 and M-NFC at two different post-bounce times.
In the outer layers of the PNS, at densities $\rho \lesssim 3\times 10^{12}$\,g\,cm$^{-3}$, the density gradient in model M-1e13 is steeper and more mass, corresponding to higher densities than in M-NFC, accumulates in the region with $3\times 10^{12}\,\mathrm{g\,cm}^{-3}\lesssim \rho\lesssim 3\times 10^{13}\,\mathrm{g\,cm}^{-3}$.
The effect is similarly, though less extremely, visible in model M-1e12.
At $\rho\gtrsim 3\times 10^{13}\,\mathrm{g\,cm}^{-3}$, however, the density profile of model M-1e13 lies lower than in model M-NFC, whereas the one of model M-1e12 is above the profile of M-NFC.
While the core of M-1e12 becomes more compact due to gravitational settling of its outer regions, the high-density interior of M-1e13 can resist the gravitational compression because of its higher reservoir of internal energy.
Again, this is fully compatible with the evolution of $\Delta E(t)$ of models M-NFC, M-1e12, and M-1e13 in Figure~\ref{fig:Energy_loss}, and the effect even amplifies at later times.
At $t_\mathrm{pb}=500$\,ms, $\Delta E$ of model M-1e13 is lower than that of the other models by about $5\times 10^{51}$\,erg, and in the more extreme model M-1e14 this gap is even $\sim$$8\times 10^{51}$\,erg at the same time.
These energy differences can be traced back to differences in the gravitational binding energies of the PNSs.

For a PNS with a lower mass ($\sim$1.35\,M$_\odot$), born in the collapse of a 9\,M$_\odot$ progenitor, we found that the decelerated energy transport out of the high-density PNS interior does not only lead to a slower contraction of the PNS core but can cause also a transient phase where the PNS radius effectively stays constant.
This has repercussions on the evolution of the neutrino luminosities and of the CCSN shock, too.
We will report these results in detail in a subsequent paper.

A curious, oscillatory density feature appears transiently in the region where $\rho \approx \rho_\mathrm{c}$ in model M-1e13 (see Figure~\ref{fig:Den_profile}; in Figure~\ref{fig:massshell} this feature can be witnessed as wiggles in the isodensity contour for $\rho_\mathrm{c} = 10^{13}$\,g\,cm$^{-3}$).
It develops after some 10\,ms post bounce during a phase when the PNS rapidly accumulates mass by accretion with a high rate and begins to quickly settle in its gravitational potential due to this mass infall. 
It disappears again at $\sim$100\,ms after bounce when the mass accretion rate has dropped and the settling of the PNS begins to slow down.

The feature has quite a complex explanation.
It is connected to the \XtoE conversion, which leads to local heating near $\rho \lesssim \rho_\mathrm{c}$ because of the rapid absorption of the conversion-produced $\nue$ and $\nuebar$.
The rising temperature causes the density to drop, since the local pressure is determined by hydrostatic equilibrium.
Therefore a trough in the density profile forms initially.
Because of the rapid accretion of mass, the density at a slightly larger radius starts to grow, exceeds $\rho_\mathrm{c}$, and \XtoE conversion stops.
Instead, the locally created $\nux$ begin to diffuse outward, and when reaching densities below $\rho_\mathrm{c}$ a part of them is again converted to $\nue$ and $\nuebar$.
The same sequence of events sets in once again, and another density trough forms.
Thus a series of density valleys and peaks emerges, appearing as a damped wave pattern on the density profile.
In this process gravitational binding energy is tapped by the creation of $\nux$, which diffuse outward, are converted to electron-type neutrinos, which are absorbed and raise the local temperature instead of transporting the accretion energy out of the PNS.
When the PNS settling slows down, the $\nux$ transport becomes sufficiently fast to smoothen the temperature profile again.
The wiggles in the density profile therefore disappear, but because of the ongoing \XtoE conversion the interior of the PNS in model M-1e13 can still not cool as fast as in model M-NFC or in the other simulations with $\rho_\mathrm{c} \lesssim 10^{12}$\,g\,cm$^{-3}$.
In model M-1e14 the oscillation feature does not develop (only a small dip in the density profile is visible) because both the gravitational settling of the PNS and the $\nux$ diffusion near the flavor conversion threshold of $\rho_\mathrm{c} = 10^{14}$\,g\,cm$^{-3}$ are too slow.
The effect also does not occur for $\rho_\mathrm{c} \lesssim 10^{12}$\,g\,cm$^{-3}$, because in these simulations the \XtoE conversion does not play a dominant role.

\section{Discussion and outlook}
\label{sec:conclusion}

We have studied the impact of FFCs on CCSN evolution and the associated neutrino emission by 1D simulations of a $20\Msol$ progenitor model.
For this purpose we have implemented FFCs in a schematic and parametric manner.
We assumed that they take place in regions with densities lower than a systematically varied threshold value.
This allows us to single out the effects when FFCs happen in different spatial domains of the CCSN.
Moreover, we assumed that FFCs immediately lead to a flavor equilibrium respecting lepton-number conservation of all neutrino flavors, in particular also ELN conservation.
Overall, our results imply that FFCs occurring in the postshock region tend to hamper SN explosions in 1D simulations.
In particular, neutrino flavor conversions deep below the shock accelerate the cooling of the neutrinospheric layers of the PNS and thus trigger a faster contraction of the PNS radius.
This, in turn, causes a faster retraction of the SN shock.
Therefore none of our 1D models shows favorable conditions for an explosion.

The exact impact of FFCs on the SN dynamics and the PNS structure can involve fairly complex feedback effects, depending on the phase of the evolution, the region where flavor conversions occur, and the type of FFCs.
Flavor conversions exterior to the SN shock do not have any noticeable impact on the SN dynamics but just modify the properties of the radiated neutrinos.
While \EtoX conversions in the bulk of the energy spectra around the spectral peaks increase the $\nux$ luminosity at the expense of the $\nue$ and $\nuebar$ luminosities, \XtoE conversions of high-energy $\nux$ in the spectral tail harden the $\nue$ and $\nuebar$ spectra.
FFCs in the postshock heating layer have the same effect on the emission properties of the neutrinos and thus lead to enhanced energy deposition through $\nue$ and $\nuebar$ absorption behind the stalled shock.
In contrast, \EtoX conversions in the cooling layer below the gain radius as well as in the region around the neutrinospheres facilitate a faster cooling of the SN environment and therefore a more rapid PNS contraction with the mentioned negative effect on the shock expansion.
If FFCs take place deeper inside the PNS, i.e., also at densities above $\sim$$10^{12}$\,g\,cm$^{-3}$, the dominant channel is \XtoE conversions because of the large $\nue$ degeneracy and corresponding underabundance of $\nuebar$ in the high-density regions.
In this case the energy release from the PNS core will be decelerated because of the important role of $\nux$ for the energy transport in the dense PNS interior.
Hence, such simulations exhibit the most pronounced enhancement of the emission of $\nue$ and $\nux$ during the cooling and deleptonization of the PNS accretion mantle in the early post-bounce phase.
But they also show the most extreme reduction of the $\nue$ and $\nuebar$ luminosities at later times when the energy diffusion out of the PNS core makes a more important contribution to the neutrino emission.
The reduction of the $\nux$ population in the PNS interior by \XtoE conversions also leads to local heating as a consequence of the rapid absorption of the electron-flavor neutrinos made by the conversions.
This prevents the PNS core from contracting as quickly as in the model without flavor conversions or in models with FFCs only at densities below $\sim$$10^{12}$\,g\,cm$^{-3}$.

All of our simulations with FFCs show a significant reduction of the numbers of emitted $\nue$ and $\nuebar$ and a corresponding enhancement of the $\nux$ emission.
Moreover, during the first 200--300\,ms after core bounce in our non-exploding models, FFCs lower the mean energy of the radiated $\nux$ compared to the simulation without FFCs, and they considerably increase the mean energies of $\nue$ and $\nuebar$, especially by lifting their high-energy spectral tails.
Remarkably, when FFCs change $\nue$ and $\nuebar$ to $\nux$, and only in that case, the radiated luminosity and mean energy of the $\nux$ also exhibit the characteristic step-like decline when the Si/O composition interface falls through the shock and the mass accretion rate onto the PNS drops steeply.
It is an interesting question deserving further study, whether such features could be unambiguous observational indicators of FFCs (or any other physics that leads to flavor equilibration in a similar manner), despite the matter resonances in the SN and earth and vacuum oscillations on the way in between.

Although our study was motivated by FFCs, the effects we found could be relevant also for slow flavor conversion modes, provided that they occur close enough to the PNS and also on scales shorter than the collision scales in this environment.
In this context it is worth noting that our prescription of flavor equilibration is indeed expected to apply for slow modes~\cite{Esteban-Pretel:2007jwl,Shalgar:2022rjj, Shalgar:2022lvv}.

Major caveats of our study are its constraints to a single stellar progenitor and to spherical symmetry.
Given the complexity and strong time and space dependence of the discussed feedback effects of FFCs on the SN evolution, it will be important to investigate other progenitors, too, in particular also those with a faster decline of the mass accretion rate and an accordingly larger gain layer, which would bring those models closer to the possibility of an explosion.
Under such circumstances FFCs in the postshock region could turn out to be more supportive for a successful shock revival.
Naturally, in order to obtain conclusive results with respect to the impact of FFCs on the SN blast, multi-dimensional simulations are indispensable.
Nonradial hydrodynamic instabilities in the postshock region in combination with neutrino heating weaken the correlation between the contracting PNS radius and the shock radius, which has a decisive influence on the shock behavior in 1D simulations.
If a larger mass in the gain layer, facilitated by the stabilizing effects of nonradial flows, meets a higher specific neutrino heating rate due to FFCs, the situation may be fundamentally different from the spherically symmetric case: instead of following the PNS contraction in 1D, the shock might be pushed outward, possibly meeting the threshold to an explosion.
Moreover, PNS convection in multi-dimensional models might have a bearing on the way how the PNS reacts to FFCs in its interior, because convective mass motions moderate the relevance of $\nux$ diffusion for the energy transport in the regions of higher density.
For all these reasons we refrain from drawing far-reaching conclusions from our 1D study on the consequences of FFCs for the neutrino-heating mechanism and explosion behavior of CCSNe.

We also emphasize that in this work we did not explicitly account for any effects of the recently unearthed phenomenon of collision induced flavor conversions \cite{Johns:2021qby, Johns:2022yqy, Xiong:2022vsy, Xiong:2022zqz, Lin:2022dek}.
Indeed, our present goal was to provide a first investigation of the impact of FFCs on dynamical CCSN simulations, disentangling FFCs from other flavor conversion phenomena.
In addition, one should note that collision induced conversions occur on scales much larger than those of FFC, implying that on small scales they cannot compete with FFC.
This makes a more dedicated exploration of such a phenomenon necessary, especially since it does not respect ELN conservation.
Should, however, collision induced conversions trigger FFCs, the overall consequences might also be mimicked by our schematic treatment.
(In this context, notice that our prescription implements a maximal effect by any flavor conversion process under the assumption of ELN conservation.)
Hence, the inclusion of collision induced flavor conversions in dynamical CCSN simulations and their impact on the physics of these settings remain open questions at this moment.
We refer the reader to Ref.~\cite{Xiong:2022vsy} for a discussion of their impact on the SN physics based on static backgrounds adopted from a CCSN simulation at different post-bounce times.

Our systematic study, based on a parametric approach in 1D, employed simple assumptions of where and how FFCs take place in CCSNe.
By this first investigation of FFC effects in hydrodynamic models, we have shown that FFCs can lead to a variety of different feedback effects on the SN dynamics, structure, and neutrino emission, all of which are strongly time and space dependent.
Further progress towards predictive neutrino-hydrodynamical simulations therefore requires multi-dimensional modeling and also a better understanding of the regions where FFCs occur in the multi-dimensional SN core and how the flavor mix is modified locally in the course of these conversions.

\section*{Acknowledgments}

J.E.\ is grateful to Robert Glas and Oliver Just for their introduction to the use of the \textsc{Alcar} code.
This work was supported by the German Research Foundation (DFG) through the Collaborative Research Centre ``Neutrinos and Dark Matter in Astro- and Particle Physics (NDM),'' Grant SFB-1258\,--\,283604770, and under Germany’s Excellence Strategy through the Cluster of Excellence ORIGINS EXC-2094-390783311.
In Copenhagen this project received funding from the Villum Foundation (Project No.~37358) and the Danmarks Frie Forskningsfonds (Project No. 8049-00038B).
We also acknowledge the use of the following software: \textsc{Matplotlib}~\cite{Matplotlib}, \textsc{Numpy}~\cite{Numpy}, \textsc{SciPy}~\cite{SciPy}, \textsc{IPython}~\cite{IPython}.



\bibliographystyle{bibi}
\bibliography{references}

\end{document}